\documentclass{article}
\usepackage[utf8]{inputenc}
\usepackage[margin=0.7in]{geometry}
\usepackage{setspace,url,hyperref,graphicx}
\usepackage{amsmath,amsthm,amssymb,amsfonts,xcolor}
\usepackage[linesnumbered,ruled]{algorithm2e}
\usepackage{wrapfig}
\newcommand{\N}{\ensuremath{\mathbb{N}}}
\newcommand{\R}{\ensuremath{\mathbb{R}}}
\newcommand{\B}{\ensuremath{\mathbb{B}}}

\newcommand{\hide}[1]{}
\newcommand{\citep}{\cite}
\newlength{\figwidth}
\usepackage{enumitem}
\usepackage{authblk}
\usepackage{biblatex}
\setlist{nolistsep,leftmargin=*}

\title{Quantum Annealing Algorithms for Boolean Tensor Networks}
\author[1*]{Elijah Pelofske}
\author[2]{Georg Hahn}
\author[1]{Daniel O'Malley}
\author[1,3]{Hristo N.\ Djidjev}
\author[1]{Boian S.\ Alexandrov}
\affil[1]{Los Alamos National Laboratory, Los Alamos, NM 87545, USA}
\affil[2]{Harvard T.H.\ Chan School of Public Health, Boston, MA 02115, USA}
\affil[3]{Institute of Information and Communication Technologies, Bulgarian Academy of Sciences, Sofia, Bulgaria}
\affil[*]{epelofske@lanl.gov}

\begin{document}
\date{}
\maketitle

\begin{abstract}
Quantum annealers manufactured by D-Wave Systems, Inc., are computational devices capable of finding high-quality heuristic solutions of NP-hard problems. In this contribution, we explore the potential and effectiveness of such quantum annealers for computing Boolean tensor networks. Tensors offer a natural way to model high-dimensional data commonplace in many scientific fields, and representing a binary tensor as a Boolean tensor network is the task of expressing a tensor containing categorical (i.e., $\{0, 1\}$) values as a product of low dimensional binary tensors. A Boolean tensor network is computed by Boolean tensor decomposition, and it is usually not exact. The aim of such decomposition is to minimize the given distance measure between the high-dimensional input tensor and the product of lower-dimensional (usually three-dimensional) tensors and matrices representing the tensor network. In this paper, we introduce and analyze three general algorithms for Boolean tensor networks: Tucker, Tensor Train, and Hierarchical Tucker networks. The computation of a Boolean tensor network is reduced to a sequence of Boolean matrix factorizations, which we show can be expressed as a quadratic unconstrained binary optimization problem suitable for solving on a quantum annealer. By using a novel method we introduce called \textit{parallel quantum annealing}, we demonstrate that boolean tensor's with up to millions of elements can be decomposed efficiently using a DWave 2000Q quantum annealer.
\end{abstract}

\setlength{\belowcaptionskip}{-15pt}
\flushbottom
\maketitle
\thispagestyle{empty}

\section{Introduction}
\label{sec:intro}
Large-scale datasets are commonplace throughout many modern scientific disciplines, such as personalized medicine, biology, space research, or climate research. Oftentimes, the underlying fundamental processes creating the data, called latent (i.e., not directly observable), remain hidden \citep{everett2013introduction}. Extracting such latent features can reveal valuable information about hidden causality and previously unknown mechanisms and relations. Usually, the high-dimensional data we observe (or generate) is sparse and stems from a lower dimensional latent space, which allows us to reduce the dimensionality (and size) of the data.

Factor analysis \citep{spearman1961general} is among the most efficient methods for educing latent (hidden) features. In the two dimensional case, the task is to approximate some data matrix $X \in \R^{n \times m}$ as a product $X \approx AB$ of two factor matrices, where $A \in \R^{n \times k}$, $B \in \R^{k \times m}$ where $k \ll n,m$ is the \textit{rank} of the factorization, and $k,m,n \in \N$. Depending on the constraints imposed upon this decomposition, different types of factorization are obtained.

For instance, imposing orthogonality on the factors results in the well known singular value decomposition (SVD) \citep{Stewart1993}, while nonnegativity leads to non-negative matrix factorization (NMF) \citep{lee1999learning}. In a lot of applications, the variables are simple dichotomies, that is \{\textit{false},\textit{true}\}, and the data contains only binary values, $\{0, 1\}$. For example, in relational databases, an object--attribute relation is represented by a Boolean variable, which takes value $1$ (\textit{true}), if the object has the attribute, or $0$ (\textit{false}), otherwise. In this case, we need to change the constraints on the factor matrices, that is, all the values of the factors have to be $0$ or $1$ as well as we need to go from simple arithmetic to Boolean algebra: the "plus" and "times" operations become the logical operations "or" and "and", respectively, which results in a Boolean matrix product \citep{miettinen2008discrete}.

\textit{Boolean matrix factorization} is a special case of factor analysis whereby the input data are given as a matrix $X \in \B^{n \times m}$, where $\B=\{0,1\}$. The task is to decompose $X = AB$ into two binary matrices $A \in \B^{n \times k}$ and $B \in \B^{k \times m}$, where $x_{ij}=\vee_{l=1}^k a_{il}y_{lj} \in \B$ and $\vee$ is the logical "or" operation ($1+1=1$). The smallest integer $k$ for which an exact representation in the form of $X=AB$ exists is called the \textit{Boolean rank} of $X$.

\textit{Tensor factorization} is the high-dimensional generalization of matrix factorization \citep{kolda2009tensor}. \textit{Tensors} (matrices of dimension three or higher) offer a natural way to represent the high-dimensional data ubiquitous in many modern scientific disciplines. The classical tensor factorization techniques such as Tucker decomposition (TD) \citep{tucker1966some} and Canonical Polyadic Decompositions (CPD) \citep{hitchcock1927expression}, with various constraints, can extract latent structures, which allow for a new type of feature extraction in high-dimensional data. CPD allows for a representation with the smallest number of parameters, but it is an NP-hard problem \citep{haastad1990tensor} that can be ill-posed \citep{de2008tensor}. TD is not feasible for high-dimensional tensors, since the memory requirement and the number of operations grow exponentially with the tensor dimension \citep{oseledets2009new}. Therefore, \textit{tensor networks}, which originated from quantum physics \citep{penrose1971applications, feynman1986quantum}, have been introduced as low-rank approximation methods for high-dimensional tensors \citep{oseledets2009breaking, oseledets2011tensor}. 

In this contribution, we consider \textit{Boolean tensor networks}, a subbranch of tensor factorization that aims to decompose tensors with binary entries into Boolean products of smaller binary tensors and matrices. Despite the problem's importance, it has been largely overlooked by computer scientists.
Related Boolean tensor factorization methods, such as CPD and TD, have been studied previously \citep{miettinen2011boolean,tucker1}, while Boolean tensor networks, have been considered in quantum physics \citep{biamonte2011categorical,biamonte2019lectures}.
We propose three algorithms: a Tucker decomposition algorithm (which comes as an iterative and recursive variant), a Tensor Train (TT) algorithm (again as iterative and recursive variant), and a Hierarchical Tucker algorithm. All three algorithms decompose an input tensor in a tree-like fashion. At their core, all three algorithms rely on solving the problem of Boolean matrix factorization, which is an NP-hard problem and the most computationally expensive step. 

In our approach, we solve the Boolean matrix factorization problem on a quantum annealer, which seems uniquely suited for this type of hard optimizations problems.
For this end, we show that the task of factoring a Boolean matrix can be expressed as a minimization of a \textit{higher order binary optimization problem (HUBO)}. A HUBO is a higher-order generalization of quadratic unconstrained binary optimization (QUBO), and its minimization is NP-hard. Importantly, for arbitrary order greater than two, HUBO can be transformed into an equivalent quadratic unconstrained binary optimization problem with (at most) a polynomial increase in the number of variables \citep{Boros2002}.

Quadratic unconstrained binary optimization, on the other hand, is the type of problem the D-Wave's quantum annealer \citep{mcgeoch2019practical, Das2008, FINNILA1994343} is designed to solve. Many important NP-hard graph problems such as maximum clique, minimum vertex cover, graph partitioning and maximum cut can be easily converted into QUBOs and solved on D-Wave \citep{ushijima2017graph, maximum_clique_journal, large_vertex_cover, junger2021quantum, Lucas2014}. Using the D-Wave 2000Q device situated at Los Alamos National Laboratory, which is employed for all the experiments presented in this article, we aim to show that quantum annealing offers a viable tool for solving large Boolean tensor network problems. In contrast to the present work, past research on utilizing quantum annealing for matrix factorization has focused mostly on non-negative matrix factorization \citep{o2018nonnegative, golden2021reverse}.

Our contribution is threefold: First, we present novel recursive algorithms for Hierarchical Tucker, Tucker, and Tensor Train networks suitable for quantum annealers, which can also be applied to other types of Boolean and non-Boolean tensor networks. These algorithms complement their iterative counterpart ready published in the literature~\cite{oseledets2009breaking}. Moreover, while the classical iterative versions were known previously, the quantum versions of both the iterative and recursive algorithms are original work of this contribution. Second, for solving Boolean matrix factorization on the quantum annealer, we design an algorithm whose required number of qubits depends only on the matrix rank, rather than its dimensions, and thereby allows tensors of very large dimensions and more than a million elements to be solved on current generation of quantum annealers as long as the tensor rank is small. In contrast, most current implementations of quantum annealer algorithms can solve problems of sizes less than $100$. Third, we apply a parallel embedding technique introduced in the literature \cite{Pelofske2022} to the tensor factorization problem, thus allowing us to solve a large number of low rank problems in parallel.

This article is a journal version and substantial extension of a published conference paper \cite{lssc}, where only the algorithm for Hierarchical Tucker factorization was introduced and no on-chip parallelism was used.

The article is structured as follows. Section~\ref{sec:methods} starts by introducing some basic notions of quantum annealing (Section~\ref{sec:basics}) and the Boolean matrix factorization algorithm which is at the heart of the tensor decompositions (Section~\ref{sec:matrix_factorization}), after which we describe the three algorithms allowing us to recursively decompose tensors into a series of lower-order tensors (Section~\ref{sec:tensor_algorithms}). Section~\ref{sec:implementation} details how we make use of the D-Wave 2000Q annealer (Section~\ref{sec:dwave}), and, in particular, how we solve low rank problems in parallel (Section~\ref{sec:parallelism}). Results from a series of experiments on random input tensors is presented in Section~\ref{sec:experiments}. The article concludes with a discussion in Section~\ref{sec:discussion}. The algorithms presented in this article have been implemented in Python and made available on a Github repository \citep{Pelofske2021_pyQBTNs}.
\section{Methods}
\label{sec:methods}
This section starts with a basic overview of quantum annealing (Section~\ref{sec:basics}). We proceed by introducing a method for Boolean matrix factorization that reformulates the factorization problem into a problem solvable on the D-Wave 2000Q quantum annealer (Section~\ref{sec:matrix_factorization}). That algorithm forms the basis of our tensor factorization algorithms, as it can be used to decompose any Boolean tensor into a Boolean tensor network using quantum annealing.

The Boolean matrix factorization algorithm consists of several phases reducing  the current problem type into a simpler one:
\begin{itemize}
    \item Boolean matrix factorization $\rightarrow$ Boolean matrix equation;
    \item Boolean matrix equation $\rightarrow$ Boolean vector equation;
    \item Boolean vector equation  $\rightarrow$ HUBO problem;
    \item HUBO problem $\rightarrow$ QUBO problem;
    \item QUBO problem $\rightarrow$ quantum annealing.
\end{itemize}

We illustrate this concept on three important tensor networks discussed in Section~\ref{sec:tensor_algorithms}.

\subsection{Basics of quantum annealing}
\label{sec:basics}
As briefly outlined in Section~\ref{sec:intro}, all of the tensor network algorithms of Section~\ref{sec:tensor_algorithms} reduce the problem of tensor factorization to the one of minimizing a quadratic unconstrained binary optimization (QUBO) problem, a task which is NP-hard. We attempt this with the help of the D-Wave 2000Q quantum annealer, manufactured by D-Wave Systems, Inc., which is briefly introduced in this section.

The quantum annealers of D-Wave Systems, Inc., are hardware devices designed to compute high quality solutions of NP-hard problems that can be expressed as the minimization of the following function,
\begin{equation}
    H(x_1,\ldots,x_{\bar{n}}) = \sum_{i=1}^{\bar{n}} h_i x_i + \sum_{i<j} J_{ij} x_i x_j,
    \label{eq:hamiltonian}
\end{equation}
where $h_i \in \R$ and $J_{ij} \in \R$ are user-specified weights that define the problem under investigation. The unknown variables $x_1,\ldots,x_{\bar{n}}$ take only two values (states). If all $x_i \in \{0,1\}$ then eq.~\eqref{eq:hamiltonian} is called a QUBO (quadratic unconstrained binary optimization) problem, and if all $x_i \in \{-1,+1\}$, then eq.~\eqref{eq:hamiltonian} is called an Ising problem. Both the QUBO and Ising formulations are equivalent \citep{maximum_clique_journal}. Many important NP-hard problems can be expressed as the minimization of eq.~\eqref{eq:hamiltonian}, see \cite{Lucas2014}.

D-Wave quantum annealers attempt to minimize eq.~\eqref{eq:hamiltonian} by mapping each of the logical variables $x_i$ to one or more physical qubits on the D-Wave quantum chip. During annealing, the \textit{Hamiltonian} operator specifies the evolution of the quantum system from the equal superposition of all qubit states to a state that corresponds to low energy solutions of eq.~\eqref{eq:hamiltonian}. This evolution of the quantum system can be described by:
\begin{equation}
H(s)=-\frac{A(s)}{2}\sum_{i=1}^n \sigma^x_i +\frac{B(s)}{2} \left( \sum_{i=1}^n h_i\sigma^z_i + \sum_{i\leq j} J_{ij} \sigma^z_i \sigma^z_j \right),
\end{equation}
where the first term encodes an equal superposition of all states. The function to be minimized, given by eq.~\ref{eq:hamiltonian}, is encoded in the second term. The dynamics with which the system transitions from the initial equal superposition, in which all bitstring solutions are equally likely, to the solution of eq.~\eqref{eq:hamiltonian} is specified through the so-called \textit{anneal path}. The anneal path is given by two functions $A(s)$ and $B(s)$ indexed by a parameter $s \in [0,1]$, called the \textit{anneal fraction}. At the start of the anneal, we have $s=0$ and $B(s)=0$, meaning that all weight is on the initial superposition. Accordingly, at the end of the anneal, we have $s=1$ and $A(s)=0$, meaning that the quantum system has fully transitioned to the problem of eq.~\eqref{eq:hamiltonian} to be solved. The main idea of adiabatic quantum annealing lays in the fact that if the aforementioned transition is performed slowly enough, the system will evolve to a solution of eq.~\eqref{eq:hamiltonian} while always staying in the ground state \cite{Albash2018, Hauke2020}.

The function given in eq.~\eqref{eq:hamiltonian} has monomials of maximal degree two, hence the ``quadratic'' in QUBO. However, in many applications, one needs to minimize functions similar to eq.~\eqref{eq:hamiltonian}, where the degrees of the monomials can be higher than two, in which case we speak of higher order binary optimization (HUBO). Conversion of a HUBO of any order larger than two into a QUBO (having only monomials of degree at most two) is always possible, and supported in the D-Wave API \cite{dwave_github}.

Experiment figures in this article were generated using Matplotlib \cite{Hunter:2007, thomas_a_caswell_2021_5773480}. 

\subsection{Boolean matrix factorization}
\label{sec:matrix_factorization}
The proposed idea of reformulating Boolean matrix factorization as a quadratic unconstrained optimization problem solvable on D-Wave consists of several problem reduction steps as follows. 

\subsubsection{From Boolean matrix factorization to Boolean matrix equation}
We consider the task of factoring $M \in \B^{\tilde{n} \times \tilde{m}}$ as the product $M=A \cdot B$ of two Boolean matrices $A$ and $B$. This is done by iteratively solving
\begin{align}
    A = \arg\min_Y d(M,Y\!B),\\
    B = \arg\min_Y d(M,AY),
    \label{eq:factorization_iteration}
\end{align}
where $d(\cdot,\cdot)$ denotes the Hamming distance. The initial values we employ for $A$ and $B$  vary depending on the tensor decomposition algorithm. Amongst others, we use the output of a non-negative SVD (NNSVD), with factors converted to Boolean factors via thresholding, as initial values \citep{nnsvd1,nnsvd2}, or initialize $A$ and $B$ with randomly generated Boolean entries. The precise choice is given in Section~\ref{sec:tensor_algorithms}. 

\subsubsection{From Boolean matrix equation to Boolean vector equation}
After noting that both aforementioned minimizations \eqref{eq:factorization_iteration} can be accomplished with the same subroutine after taking transposes, we are looking at the problem $B = \arg\min_Y d(M,AY)$.
Next, the latter equation can be decomposed into a set of independent column-wise equations, leading to
\begin{equation}
    B_i = \arg\min_{y} d(M_i,Ay)~\text{for}~i \in \{ 1,\ldots,\tilde{m} \},\label{eq:bmt2bme}
\end{equation}
where $M_i$ denotes the $i$-th column of matrix $M$ and $y=Y_i$. Next we show how such type of Boolean equation can be further reduced into a HUBO and then to a QUBO problem suitable for a quantum annealer.

\subsubsection{From Boolean vector equation to HUBO}
To solve eq. \eqref{eq:bmt2bme}, denote the set of all indices with entry \textit{true} in column $M_i$ as $T_i = \{j: M_{ji}=1 \}$, and the set of all indices with entry \textit{false} in column $M_i$ as $F_i = \{j: M_{ji}=0 \}$. The Hamming distance $d(M_i,Ay)$ can then be expressed as
\begin{equation}
    d(M_i,Ay) = C - \sum_{j \in T_i} f( (A^\top)_j \odot y) + \sum_{j \in F_i} f( (A^\top)_j \odot y),
    \label{eq:HUBO}
\end{equation}
where the number of non-zero entries in column $M_i$ is a constant $C$, the symbol $\odot$ denotes an entrywise multiplication of two vectors, and $f(x_1,\ldots,x_{\tilde{n}}) = 1 - \prod_{i=1}^{\tilde{n}} (1-x_i)$. Importantly, since $y_1,\ldots,y_m$ are unknown, eq.~\eqref{eq:HUBO} becomes a higher order polynomial in binary variables $y_i$, thus making eq.~\eqref{eq:HUBO} a HUBO problem.

\subsubsection{From HUBO to QUBO}
In a HUBO, there are monomials of degree greater than two, e.g., $x_1x_2x_3$. One way to convert a HUBO into a QUBO is to convert each monomial into a quadratic polynomial by introducing auxiliary variables, e.g., $u_{12}=x_1x_2$, which are substituted into the monomial, thereby reducing its degree. As mentioned in Section~\ref{sec:basics}, in our implementation we employ features included in the D-Wave API for converting the HUBO of eq.~\eqref{eq:HUBO} into a QUBO, in order to be able to solve it with the D-Wave annealer. Further details of the D-Wave implementation are given in Section~\ref{sec:implementation}.

\vspace{0.2cm}
%\begin{wrapfigure}{l}{0.5\textwidth}
\begin{algorithm}%[H]
    \caption{\texttt{column\_factorization}}
    \label{algo:column_factorization}
    \SetKwInOut{Input}{input}
    \SetKwFor{Loop}{repeat}{}{end}
    \Input{matrix $M$, integer $N$, initial state matrix $A$, number of anneals $n_A$, global list $T$ of precomputed solutions}
    $r \leftarrow$ number of columns  of $M$\\
    $B \leftarrow [~]$ (empty matrix)\\
    \For{$i \in \{1,\ldots,r\}$}{
        Compute HUBO $H$ according to eq.~\eqref{eq:HUBO} for column $M_i$, initial state $A$, and rank $r$\\
        Convert $H$ to QUBO $Q$ using the D-Wave API with penalty (strength) set to the maximum of the absolute value of any coefficient in $H$\\
        \If{$Q=\emptyset$}{$s \leftarrow$ random vector of $0$ and $1$ having a length equal to the number of variables in $Q$}
        \ElseIf{$Q \in T$}{Look up the known solution $s$ of $Q$ in $T$}
        \Else{Call quantum annealer for $Q$ and return $n_A$ anneals in set $D$\\
            Obtain best annealing solution $s$ from $D$ after majority vote post post-processing
            \\
            Add the tuple $(Q,s)$ to $T$}
        Add $s$ as new column to matrix $B$\\
    }
    \Return $B$
\end{algorithm}
%\end{wrapfigure}

\subsubsection{Algorithmic details}
The complete Boolean matrix factorization algorithm is summarized in Algorithm~\ref{algo:matrix_factorization}.

It relies on Algorithm~\ref{algo:column_factorization}, which formalizes the column-wise iterative factorization method of eq.~\eqref{eq:factorization_iteration}. For each column $M_i$ ($i \in \{1,\ldots,r\}$) of a given matrix $M$ which is to be factored into a product $AB$, we construct the HUBO expressing the distance between $M_i$ and the corresponding column in the factorization $AB$. After converting the HUBO into a QUBO $Q$, three cases are considered in preparation for solving $Q$ on D-Wave. If the QUBO is "empty" (i.e., only has zero coefficients), the solution is set to a random bitstring of appropriate length. Otherwise, to save computational time, we look up if $Q$ has been solved in a another problem previously. For this, a global list $T$ is utilized. If so, we look up the solution, otherwise we minimize $Q$ with a D-Wave call and add the best solution to $T$. For each sample returned by the D-Wave call, we post-process the solution using majority vote. Post-processing is a necessary step in the case of \textit{broken chains}, meaning that an embedded chain (represented by linked physical qubits) disagree about the state of the logical variable (i.e. physical qubits in a chain take values of both $0$ and $1$). Algorithm~\ref{algo:column_factorization} returns a matrix with $r$ columns, one for each column in $M$. Each column $i$ contains the QUBO solution (factorization) of $M_i$.

%\noindent
%\begin{minipage}{0.38\textwidth}
\begin{algorithm}[H]%[t]
    \caption{\texttt{iterative\_matrix\_factorization}}
    \label{algo:single_iter_matrix_factorization}
    \SetKwInOut{Input}{input}
    \SetKwFor{Loop}{repeat}{}{end}
    \Input{matrix $M$, initial state Boolean matrices $A$ and $B$, maximum number of converged minima iterations $L_c$, maximum number of iterations $L_h$ }
    \If{$M=A \cdot B$}{\Return $A, B$}
    $i \leftarrow 0$\\
    \While{ \textnormal{number of repeated minima in the last} $L_c$ \textnormal{iterations} $\leq L_c$}{
        $i \leftarrow i+1$\;
        \If{$i > L_h$}{\Return minimum-error solution $A, B$}
        $B \leftarrow$ \texttt{column\_factorization}($M$, ncol($M$), $A$, $B$)\\
        \If{$M=A \cdot B$}{\Return minimum-error solution $A, B$}
        $A \leftarrow$ \texttt{column\_factorization}($M^\top$, nrow($M^\top$), $B^\top$, $A^\top$)\\
        \If{$M=B^\top \cdot A^\top$}{\Return minimum-error solution $A^\top, B^\top$}
    }
    \Return minimum-error solution $A, B$\\
\end{algorithm}
%\end{minipage}%
%\hfill
%\begin{minipage}{0.57\textwidth}

Using Algorithm~\ref{algo:column_factorization}, we can state the full matrix factorization in Algorithm~\ref{algo:single_iter_matrix_factorization}, as follows. We start with a Boolean matrix $M$ to be factored, and two initial state Boolean matrices $A$ and $B$. If $M=A \cdot B$, where $ \cdot $ denotes the multiplication operation of two matrices, the factorization is complete and the algorithm stops. Otherwise, we alternatively solve the coupled equations of eq.~\eqref{eq:factorization_iteration} to iteratively approximate the two factors $A$ and $B$. For the single column factorization, Algorithm~\ref{algo:column_factorization} is called, where ncol$(M)$ denotes the number of columns of $M$. Both coupled equations can be solved with Algorithm~\ref{algo:column_factorization} after transposition of all matrices. The number of repeated minima found and a cutoff on the number of iterations serve as termination criteria. The termination criteria of repeated minima is implemented because it serves as an indication that the algorithm got stuck in a local minimum. The second termination criteria is used so that the algorithm is guaranteed to terminate. The algorithm returns the two factors $A$ and $B$, as well as a vector of Hamming distances.

Since Algorithm~\ref{algo:single_iter_matrix_factorization} requires suitable initial state Boolean matrices, we refine the algorithm to work without starting values. Algorithm~\ref{algo:matrix_factorization} takes as input a Boolean matrix $M$ to be factored, as well as the auxiliary parameters chosen by the user $N_\text{states}, Rand_\text{dur}, L_c, L_h$ (in our experiments we set each of these parameters to fixed constants as shown in the input of Algorithm~\ref{algo:matrix_factorization}). 
First, a non-negative SVD (NNSVD) \cite{nnsvd1} is computed, and its result serves as input to Algorithm~\ref{algo:single_iter_matrix_factorization}. In order to obtain Boolean factors from NNSVD, we rounded each element of the resulting A and B initial states to be binary (i.e. either 0 or 1). If $M=A \cdot B$ can be successfully factored, the result is returned. Otherwise, a number of $N_\text{states}$ random matrices are generated as starting values for Algorithm~\ref{algo:single_iter_matrix_factorization}. After calling Algorithm~\ref{algo:single_iter_matrix_factorization}, the results are saved, in particular the smallest Hamming distance obtained. After running those $N_\text{states}$ attempts, each attempt using a very small number of iteration denoted by $Rand_\text{dur}$, the one achieving the smallest Hamming distance is used one last time as starting point for Algorithm~\ref{algo:single_iter_matrix_factorization}, this time using the maximum allowed iteration parameters $L_c, L_h$. If any factorization successfully achieves an exact representation $M=A \cdot B$, it is returned, otherwise the Boolean matrices $A,B$ achieving the minimal Hamming distance are returned.

\begin{algorithm}[t]%[t]
    \caption{\texttt{matrix\_factorization}}
    \label{algo:matrix_factorization}
    \SetKwInOut{Input}{input}
    \SetKwFor{Loop}{repeat}{}{end}
    \Input{matrix $M$, $N_\text{states} \leftarrow 20$, $Rand_\text{dur} \leftarrow 2$, $L_c \leftarrow 10$, $L_h \leftarrow 100$}
    $T \leftarrow \emptyset$; $E \leftarrow \emptyset$\\
    $r \leftarrow \texttt{rank}(M)$\\
    $A_\text{init}, B_\text{init} \leftarrow$ \texttt{Boolean\_NNSVD}($M$, $r$)\\
    $A, B \leftarrow$ \texttt{iterative\_matrix\_factorization}($M$, $A_\text{init}$, $B_\text{init}$, $L_c$, $L_h$)\\
    \If{$M=A \cdot B$}{\Return $A,B,r$}
    \For{$n \in \{1,\ldots,N_\textnormal{states}\}$}{
        Draw $p$ from a uniform distribution in $(0.1,0.9)$\\
        $A_\text{init} \leftarrow$ random matrix in $\B^{\text{nrow}(M) \times r}$ with entry $1$ ($0$) chosen with probability $p$ ($1-p$)\\
        $B_\text{init} \leftarrow$ random matrix in $\B^{r \times \text{ncol}(M)}$ with entry $1$ ($0$) chosen with probability $p$ ($1-p$)\\
        $A, B, H \leftarrow$ \texttt{iterative\_matrix\_factorization}($M$, $A_\text{init}$, $B_\text{init}$, $L_c$, $Rand_\text{dur}$) \\
        \If{$M=A \cdot B$}{\Return $A,B,r$}
        $E \leftarrow E \cup \{(A,B)\}$\\
    }
    $(A_\text{init},B_\text{init}) \leftarrow \arg\min_{(A,B) \in E} d(M,A \cdot B)$\\
    $A, B \leftarrow$ \texttt{iterative\_matrix\_factorization}($M$, $A_\text{init}$, $ B_\text{init}$, $L_c$, $L_h$) \\
    \If{$M=A \cdot B$}{\Return $A,B,r$}
    \Else{\Return $A_\text{init},B_\text{init},r$ if $H_\text{init} \leq H$, and $A,B,r$ otherwise}
\end{algorithm}
%\end{minipage}

\subsection{From Boolean tensor networks to Boolean matrix factorization}
\label{sec:tensor_algorithms}
This section discusses our high level algorithms, i.e., the reductions of Boolean tensor networks to Boolean matrix factorization. Boolean tensor network algorithms generally consist of sequences of the following three types of operations: unfolding and reshaping, which reorder the elements of the tensor or matrix and which are described in more detail below, and Boolean matrix factorization. The first two operations can be efficiently performed on a classical computer in linear time. The third operation type, which is of a combinatorial type and NP-hard, we solve on the quantum annealer.

\begin{wrapfigure}{r}{0.49\textwidth}
    \centering
    \vspace{0.8em}
    \includegraphics[width=0.49\textwidth]{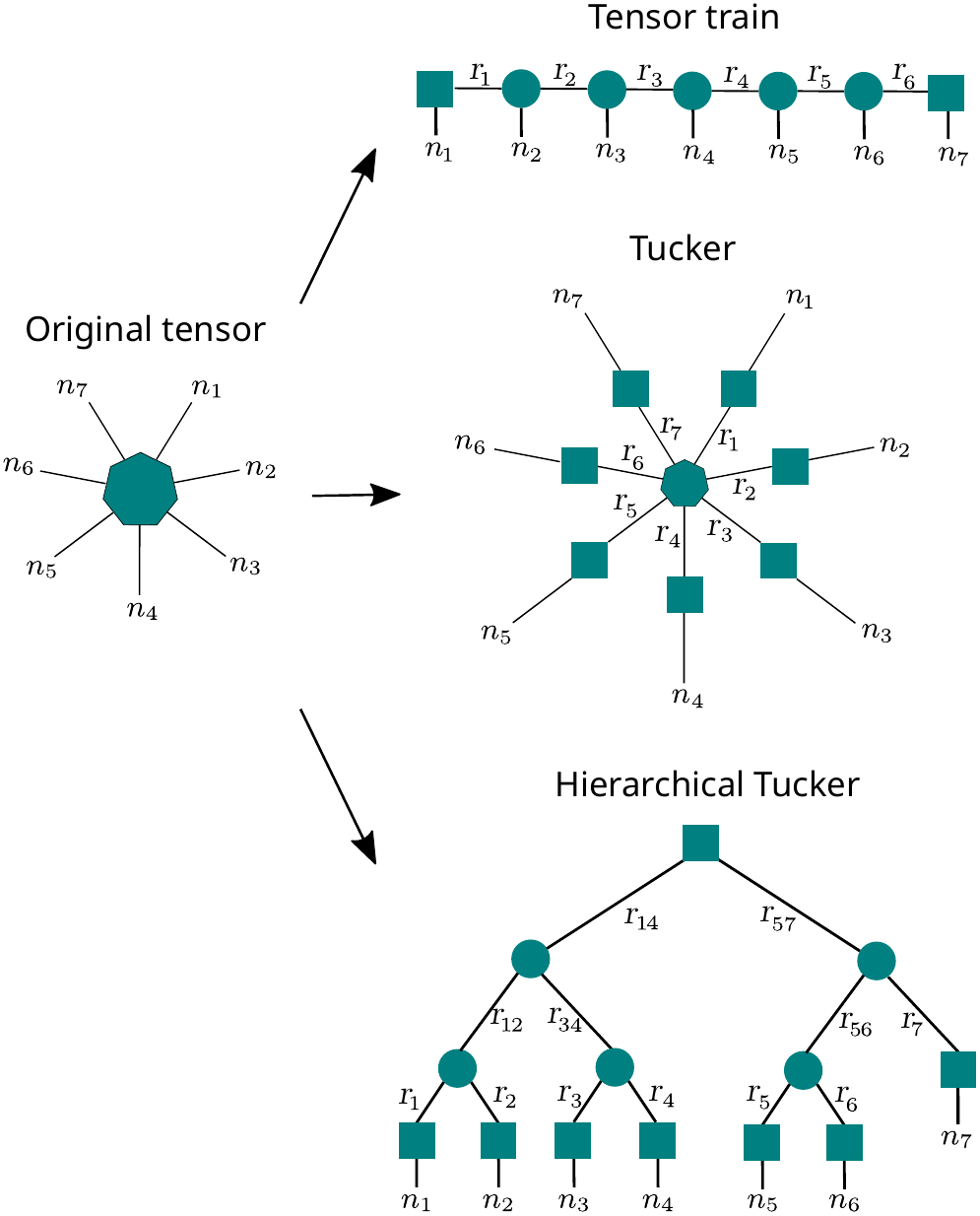}
    \caption{Decomposition of an original input tensor (left) using: Tensor train (top), Tucker (middle), and Hierarchical Tucker algorithm. A heptagon symbolizes a 7-dimensional tensor, a circle encodes a 3-dimensional tensor, and a square encodes a matrix.\label{fig:schematic} \vspace{3em}}
\end{wrapfigure}

To illustrate our approach and evaluate its efficacy and efficiency on specific problems, we use three of the most popular tensor network models. Those are the Tensor Train, Tucker, and Hierarchical Tucker networks, illustrated in Figure \ref{fig:schematic}. We then describe algorithms for constructing such networks suitable for our approach. The exact implementation of these algorithms can also be found on Github \cite{Pelofske2021_pyQBTNs}.

In all pseudocodes, we assume that our input tensor has the attributes $.order$ (which returns the order as integer) and $.dimensions$ (which returns a list of tensor dimensions, with the list length being equal to the order). Moreover, we denote with $a[:N]$ the subvector or subarray of $a$ consisting of the first $N$ elements (excluding position $N$ itself), and with $a[N:]$ the subvector or subarray of $a$ consisting of all elements from position $N$ (included) onwards. Although all of our algorithms can use multi-rank factorization, for simplicity of the comparisons in Section \ref{sec:experiments}, we assume each factorization rank is the same and given in advance, and we assume that each tensor dimension size is the same. Finding the appropriate rank value is, in general, a hard problem, and beyond the scope of this paper.

\begin{wrapfigure}{r}{0.49\textwidth}
    \vspace{0.5em}
    \centering
    \includegraphics[width=0.39\textwidth]{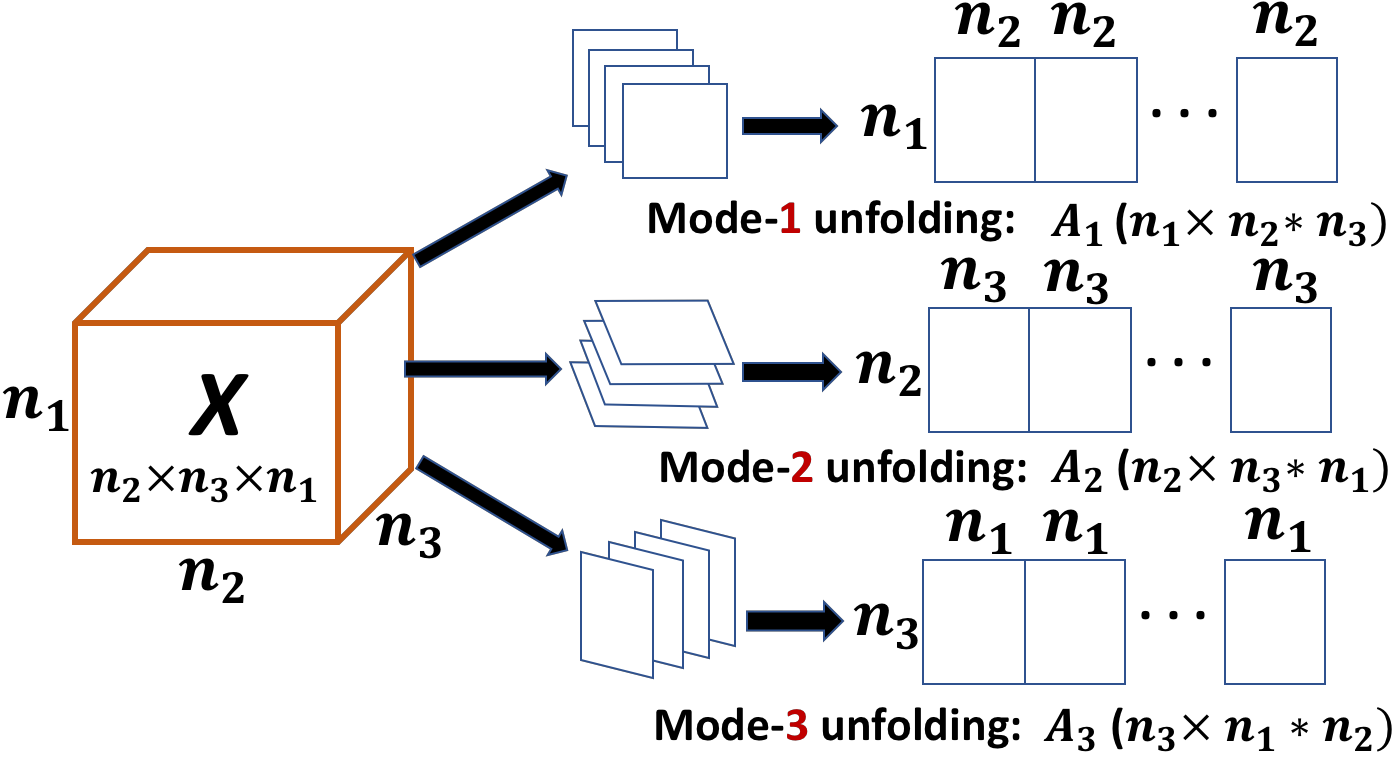}
    \caption{Visualization of the unfolding operation. A higher-dimensional input tensor $X$ (of dimension $3$ in the example) is unfolded into a matrix by "stitching" it together alongside any of the $3$ dimensions. \vspace{3em}}
    \label{fig:unfolding}
\end{wrapfigure}

\subsubsection{Basic definitions}
\label{sec:definitions}
Three basic operations occur throughout the tensor algorithms presented in the following subsections. Those operations are briefly discussed in this section.

All recursive implementations require the computation of a \textit{splitting point} (denoted with the variable $split\_point$). This is achieved with the help of a function $\texttt{split}(T)$ for an input tensor $T$, which returns the splitting point (an integer in the set $\{1,\dots,T.order\}$) and four variables denoted with $d_1$, $d_2$ and $dims_1$, $dims_2$. The quantity $d_1$ is the product of the dimensions up to $split\_point$, and $d_2$ is the product of the dimensions from $split\_point + 1$ up to $T.order$, while the lists of the corresponding dimensions are $dims_1$ and $dims_2$. The precise implementation of the function \texttt{split} differs between the three tensor algorithms we consider, and is given individually.

Moreover, our algorithms require the factorization rank of the input tensor, which we assume can be computed with a function \texttt{rank}, where the function \texttt{rank} takes a matrix (which is the unfolded tensor) as input and returns a positive integer $r$ (the rank). We do not specify further how to compute the rank for a tensor (unfolded as a matrix), as this can be a computationally hard problem. In our experiments of Section~\ref{sec:experiments}, the rank is always specified ahead of time (in order to compare the differences of factorization when using different ranks).

Finally, the unfolding operation used in Algorithms~\ref{algo:tensor_train} to ~\ref{algo:hierarchical_tucker_factorization} is visualized in Figure~\ref{fig:unfolding}. It shows that a tensor of order $3$ can be unfolded by iterating alongside any of its dimensions, and "stitching" together the slices (which are matrices in the example) to a new matrix. The unfolding operation generalizes to higher dimensions (also called matrization or fattening of a tensor), in which case it reduces the dimension by one. Recursive application allows one to reduce the dimension of any tensor until a matrix level is reached. The order in which a tensor is unfolded is not unique, thus leading to several unfolded representations. The unfolding is carried out by a function \texttt{unfold}.

Algorithms~\ref{algo:tensor_train} to ~\ref{algo:hierarchical_tucker_factorization} also rely on an operation called \texttt{reshaping}, which changes the shape (or the order) of the tensor without changing the data and number of elements. Reshaping rearranges the elements of a matrix into either matrices of other dimensions, or higher order tensors, see Figure~\ref{fig:reshape}. A precise mathematical definition of the unfolding and reshaping operations can be found in the literature \cite{kolda2009tensor}.

\subsubsection{Tensor train algorithm}
\label{sec:tensor_train}
\begin{figure}
\centering
\begin{minipage}{0.49\textwidth}
  \centering
    \includegraphics[width=0.7\textwidth]{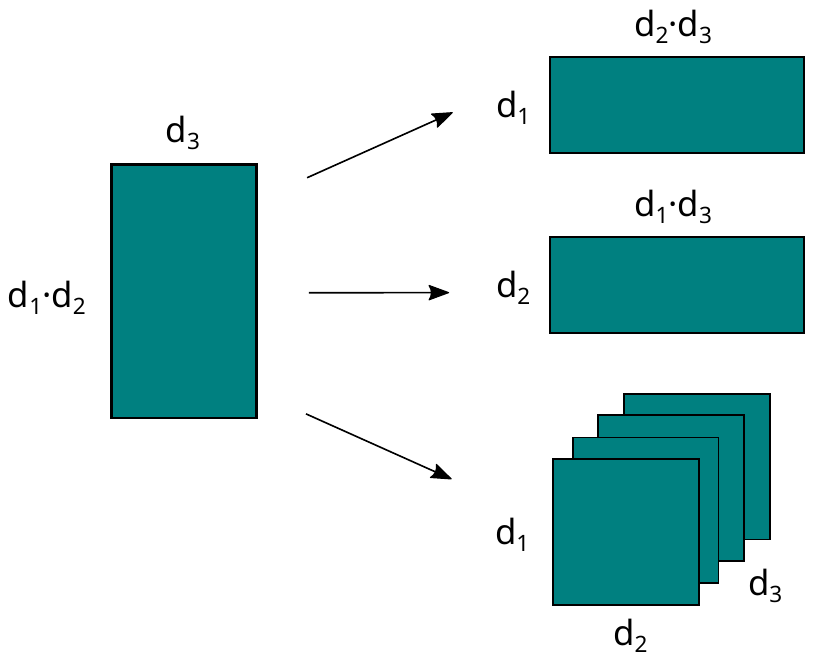}
    \caption{Visualization of the reshaping operation. A matrix of dimensions $d_1 d_2$ by $d_3$, where $d_1,d_2,d_3 \in \N$, can be reshaped into another matrix of different dimensions (right top and middle) or into am order-3 tensor (right bottom).}
    \label{fig:reshape}
\end{minipage}%
\hfill
\begin{minipage}{0.40\textwidth}
  \centering
    \centering
    \includegraphics[width=\textwidth]{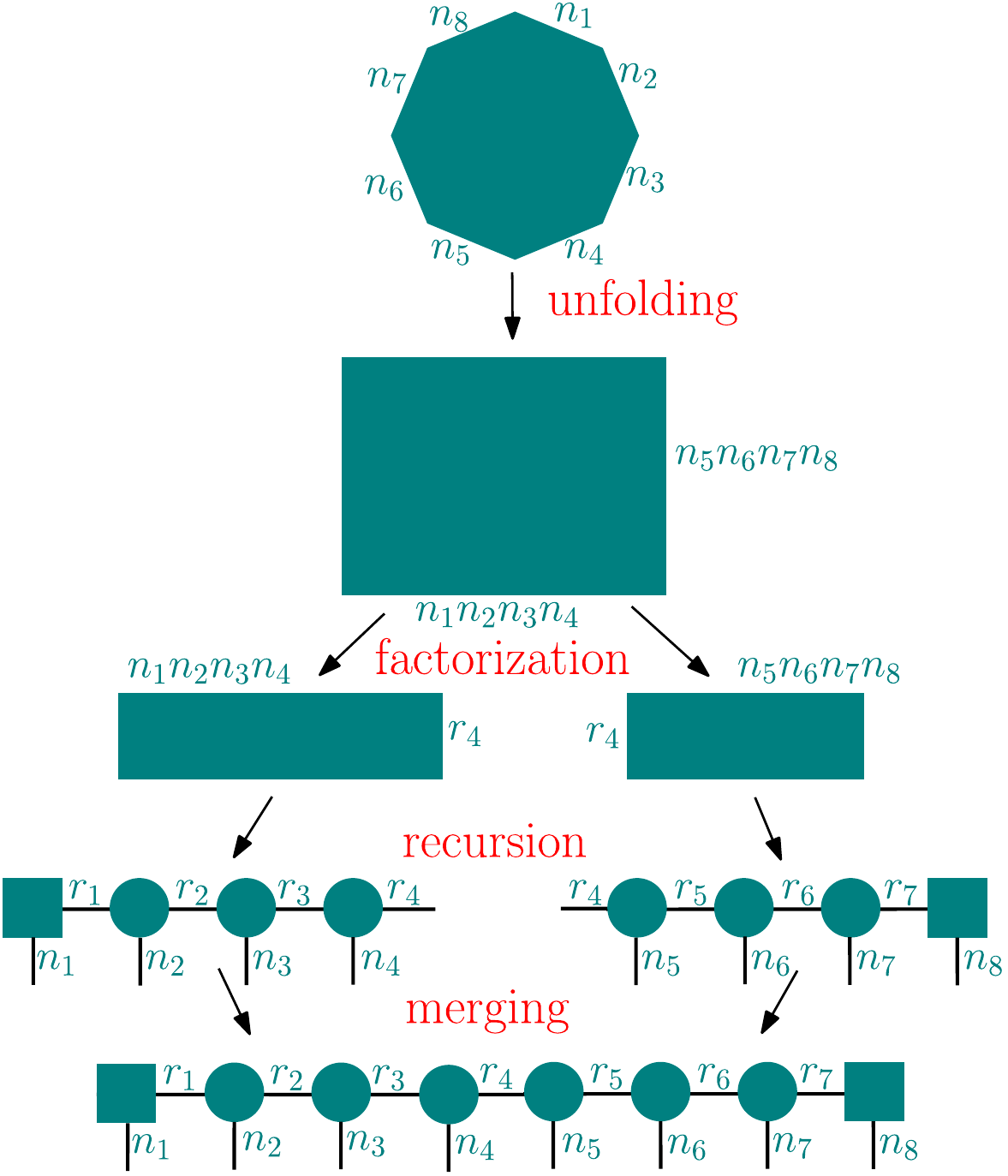}
    \caption{The phases of the recursive tensor train algorithm. The octagon is an input order-8 tensor, rectangles are matrices, and circles are order-3 tensors.}
    \label{fig:TT-diagram}
\end{minipage}
\end{figure}

A tensor train network for a tensor of order $d$ is a linear product of a matrix, $d-2$ order-3 tensors, and another matrix (Figure~\ref{fig:schematic}). We implement two versions of the tensor train algorithm, iterative and recursive. Since the iterative version has been previously described in the literature, we show the phases of the recursive algorithm, which is new, on an example order-8 tensor. In Figure~\ref{fig:TT-diagram}, the input tensor is first unfolded into a matrix, that matrix is factored as a product of two matrices, then each of those matrices is converted into a tensor-train-like structure by applying the same algorithm recursively, and finally the two parts are merged into a single tensor train network. The tensor networks produced at the intermediate levels of the recursion do not always have the structure of the tensor trains as illustrated on Figure~\ref{fig:schematic} since one or both of the matrices at its ends can be replaced by order-3 tensors (in order to make future merging or contraction possible).

Algorithm~\ref{algo:tensor_train} gives more details of this procedure. The input to the algorithm is the tensor $T$ to be factored, its rank $r$, and a parameter \textit{rec}, which determines if the tensor is being split at the midpoint of its dimension (resulting in a recursive method), or at each dimension successively (effectively resulting in an iterative method). The $split\_point$ for Algorithm~\ref{algo:tensor_train} is defined as $\lceil \left( T.order-\gamma \right)/2 \rceil$ in the recursive case, where $\gamma=1$ if all dimensions in $T$ equal the ranks in $T$ from the second one onward, or $\gamma=0$ otherwise. In the iterative case, the $split\_point$ is defined as $1 + \gamma$, where $\gamma=1$ if the first dimension of $T$ is equal to the first rank of $T$, or $\gamma=0$ otherwise. After the splitting point is computed, $T$ is reshaped into an appropriate matrix $M$ using two dimensions called $d_1$ and $d_2$ (see Section~\ref{sec:basics}). The matrix $M$ is then factored into two matrices $M_1$ and $M_2$ with the help of Algorithm~\ref{algo:column_factorization}. Afterwards, Algorithm~\ref{algo:tensor_train} is called recursively on $M_1$ (lines 4-7) and $M_2$ (lines 11-14), given the dimension of each is still large enough to allow for further decomposition. Here, $T.dimension[0]$ refers to the dimension of the first component of $T$, and $T.dimension[-1]$ refers to the dimension of the last component of $T$. Otherwise (lines 8-10 and 15-17), if the number of dimensions is three, the matrix is reshaped as an order-3 tensor, or if it is two, then it is just left as a matrix. The algorithm returns the tensor train as a list of the order-3 tensors and matrices, where $TT_1+TT_2$ denotes the concatenation of the two lists given by $TT_1$ and $TT_2$.

\subsubsection{Tucker algorithm}
\label{sec:tucker}
A Tucker network for a tensor of order $d$, dimensions $n_1,\dots,n_d$, and ranks $r_1,\dots,r_d$ is a product of an order-$d$ tensor with dimensions $r_1,\dots,r_d$ and $d$ matrices with dimensions $r_i\times n_i$, as shown on Figure~\ref{fig:schematic}. Our Tucker decomposition algorithm also comes in two flavors, an iterative and a recursive variant. Algorithm~\ref{algo:tucker_iterative} presents the iterative version. Its input consists of a tensor $T$ to be factored, and a desired rank $r$. The algorithm works by iterating through all possible orders from $1$ to $T.order$. At the $n$'th iteration, the current tensor $T_n$ is reshaped into a matrix, which is then factored into a product $M_1 \cdot M_2$ with the help of Algorithm~\ref{algo:matrix_factorization}. The first (smaller) factor $M_1$ is appended to the list of factors (initialized with the empty list at the start), which the algorithm returns upon termination. The second factor $M_2$ is reshaped appropriately again into a matrix, and subsequently factored at the next iterations.

%\begin{wrapfigure}{r}{0.49\textwidth}
%\vspace{-1.5em}
\begin{algorithm}%[H]
    \caption{\texttt{tensor\_train}}
    \label{algo:tensor_train}
    \SetKwInOut{Input}{input}
    \Input{tensor $T$, boolean $rec$ (algorithm is recursive if true, iterative if false)}
    $d_1,d_2,dims_1,dims_2, split\_point \leftarrow \texttt{split}(T, rec)$\\
    $M \leftarrow \texttt{unfold}(T, (d_1, d_2))$\\
    $M_1, M_2, r \leftarrow$ \texttt{matrix\_factorization}$(M)$\\
    \If{$split\_point > 2$ or $(split\_point > 1$ and $T.dimension[0] > r)$}{
        $T_1 \leftarrow \texttt{reshape}(M_1, dims_1)$\\
        $TT_1 \leftarrow \texttt{tensor\_train}(T_1, rec)$
    }
    \Else{
        $TT_1 \leftarrow [\texttt{reshape}(M_1, T_1.dimensions)]$
    }
    \If{$d_2 > 3$ or $(d_2 > 2$ and $T.dimension[-1] > r)$}{
        $T_2 \leftarrow \texttt{reshape}(M_2, dims_2)$\\
        $TT_2 \leftarrow \texttt{tensor\_train}(T_2, rec)$
    }
    \Else{
        $TT_2 \leftarrow [\texttt{reshape}(M_2, T_2.dimensions)]$
    }
    \Return{$TT_1 + TT_2$}
\end{algorithm}
%\end{wrapfigure}

\begin{wrapfigure}{r}{0.49\textwidth}
%\vspace{-1em}
\begin{algorithm}[H]
    \caption{\texttt{iterative\_tucker}}
    \label{algo:tucker_iterative}
    \SetKwInOut{Input}{input}
    \Input{tensor $T=T_1$}
    $matrix\_factors \leftarrow \{\}$ (empty list)\\
    \For{$n=1,2,\ldots, T.order$}{
        $reshaped\_T \leftarrow \texttt{unfold}(T_n, n)$\\
        $M_1, M_2, r \leftarrow \texttt{matrix\_factorization}(reshaped\_T)$\\
        Append $M_1$ to list $matrix\_factors$\\
        $dimension\_list \leftarrow T_n.dimensions$\\
        $dimension\_list[n] \leftarrow r$\\
        $T_{n+1} \leftarrow \texttt{reshape}(M_2, dimension\_list)$
    }
    \Return{$core, matrix\_factors$}
\end{algorithm}
\end{wrapfigure}

The Tucker decomposition algorithm can also be formulated in a recursive fashion. Details are provided in Algorithm~\ref{algo:tucker_recursive}. Its input consists of the tensor $T$ to be factored, the desired rank $r$, and a parameter $min\_rec\_ord$, defining the minimum recursive order for termination of the algorithm, which we set to $4$ in our experiments. The reason for introducing such a minimum recursive order is the fact that, upon reaching small orders, the computational cost of the recursion increases dramatically due to very high recursion levels. The minimum recursion order must be an even integer.

The algorithm works similarly to Algorithm~\ref{algo:tucker_iterative}. After setting the splitting point to $\lceil \left( T.order - \gamma \right)/2 \rceil$, where $\gamma=1$ if all dimensions in $T$ equal the ranks in $T$ from the second one onward or $\gamma=0$ otherwise, we aim to split $T$ at that point into two tensors of lower dimension. This is done as usual by reshaping into a matrix $M$, which is then factored into two factors $M_1$ and $M_2$ with the help of Algorithm~\ref{algo:matrix_factorization}. Given the splitting point is still larger than the minimal order for continuing the recursion (parameter $min\_rec\_ord$) the algorithm is called recursively for $M_1$. For $M_2$, the recursion is called if the order of $T_1$ is at least $min\_rec\_ord$. Otherwise, if the dimension does not allow for a recursive call, $M_1$ or $M_2$ are decomposed with the help of iterative Tucker (Algorithm~\ref{algo:tucker_iterative}). Algorithm~\ref{algo:tucker_recursive} returns the core and the factors of the decomposition as lists, where $factor_1+factor_2$ denotes the concatenation of the two lists given by $factor_1$ and $factor_2$.

\subsubsection{Hierarchical Tucker algorithm}
\label{sec:hierarchical_tucker}
\begin{wrapfigure}{r}{0.49\textwidth}
    \centering
    \vspace{0.1em}
    \includegraphics[width=0.32\textwidth]{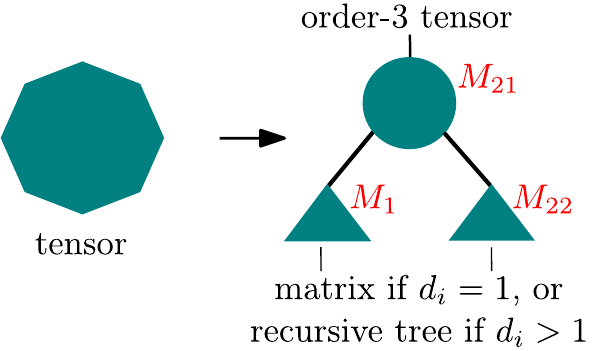}
    \caption{Schematic of the recursion in Algorithm~\ref{algo:hierarchical_tucker_factorization} for $M_1$ and $M_2$. The root of the tree (the core) is either a matrix, at the top of the recursion hierarchy, or an order-3 tensor that connects to its parent node and its two children.\vspace{0.5em}}
    \label{fig:ht}
\end{wrapfigure}

A Hierarchical Tucker network for a tensor of order $d$ is a product of a matrix, $d-2$ order-$3$ tensors, and $d$ other matrices, connected using the binary-tree pattern shown in Figure~\ref{fig:schematic}.
Our Boolean Hierarchical Tucker Network (BHTN) algorithm is a recursive one (see Figure~\ref{fig:ht}), consisting of a sequence of reshaping and matrix factorization operations.
We start with an order-$d$ input tensor $T$. The task is to transform $T$ into a BHTN $HT$, where $HT$ denotes both the BHTN and its associated decomposition tree.

Let $T(n_1,\ldots,n_s,q)$ denote the tensor at some recursion level, where $n_i$ is the size in the $i$-th dimension and $q$ is a rank used in the factorization at the higher-level recursion ($q=1$ initially). We define $s_2 = \lfloor \frac{s}{2} \rfloor$. 
Our algorithm, given as pseudocode in Algorithm~\ref{algo:hierarchical_tucker_factorization}, performs a series of reshaping and splitting operations leading to the output subtree $HT$, which is a BHTN of $T$. We begin by unfolding $T$ into a matrix $M=M(n_1,\ldots,n_{s_2},n_{s_2+1},\ldots,n_s q)$. As long as $s > 3$, the following steps are executed. 

First, using the matrix factorization algorithm of Section~\ref{sec:matrix_factorization}, we split $M$ into the product of two matrices of given dimensions, that is
\begin{align}
    M \rightarrow \enskip & M_1(n_1,\ldots,n_{s_2},r_{(1,s_2)}) \cdot M_2(r_{(1,s_2)},n_{s_2 +1},\ldots,n_s q),
    \label{eq:step1}
\end{align}
where $M_1$ and $M_2$ denote matrices containing the elements of the left and right branches (subsubtree) of the recursion (decomposition subtree $HT$), respectively, and $r_{(1,s_2)}$ is the rank of the factorization. Additionally, we will need to extract, from $M_2$, one order-3 tensor called the \textit{core}, which will be the root of $HT$ connecting the left and right branches. The core also connects $HT$ to its parent 3-d tensor.

Next, both $M_1$ and $M_2$ are prepared for further factorization using two separate recursive calls, given their orders ($d_1$ for $M_1$, $d_2$ for $M_2$) are larger than one. To be precise, the dimension $q$ is transferred from the columns to the rows of $M_2$ using the reshape operation, yielding
\begin{align}
    %\nonumber
    \eqref{eq:step1} \rightarrow  & M_1(n_1,\ldots,n_{s_2},r_{(1,s_2)}) \cdot
    M_2(q r_{(1,s_2)},n_{s_2+1},\ldots,n_s).
    \label{eq:step2}
\end{align}
Leaving $M_1$ unchanged, and extracting the core (shaped as matrix $M_{21}$) from $M_2$ yields
\begin{align}
    %\nonumber
    \eqref{eq:step2} \rightarrow  & M_1(n_1,\ldots,n_{s_2},r_{(1,s_2)}) \cdot
    %\nonumber
    M_{21}(q r_{(1,s_2)},r_{(s_2+1,s)}) \cdot
    M_{22}(r_{(s_2+1,s)},n_{s_2+1},\ldots,n_s).
    \label{eq:step3}
\end{align}
Recursively applying this decomposition to each generated subtree, as well as reshaping $M_{22}$ into an order-3 tensor, eventually yields
\begin{align}
    \eqref{eq:step3} & \rightarrow \enskip HT_\text{left}([1,s_2],r_{(1,s_2)}) \times
    T_\text{core}(q,r_{(1,s_2)},r_{(s_2 +1,s)}) \times
    HT_\text{right}([s_2+1,s],r_{(s_2+1,s)})
    = HT([1,s],q),
    \label{eq:step4}
\end{align}
where $[k_1,k_2] := \{ k_1,k_1+1,\dots,k_2 \}$. Any tensor which is flattened out as a matrix can be decomposed in this fashion so long as $s>3$. The decomposition is constructed explicitly for $s \leq 3$. Our algorithm relies on two operations only, reshaping (see Section~\ref{sec:basics}) and factorization (see Section~\ref{sec:matrix_factorization}).

\bigskip
\noindent
\begin{minipage}{0.49\textwidth}
\begin{algorithm}[H]
    \caption{\texttt{recursive\_tucker}}
    \label{algo:tucker_recursive}
    \SetKwInOut{Input}{input}
    \Input{tensor $T$, integer $min\_rec\_ord = 4$ defining the minimum order for termination of the recursion}
    $d_1,d_2,dims_1,dims_2, split\_point \leftarrow \texttt{split}(T)$\\
    $M \leftarrow \texttt{unfold}(T, (d1, d2))$\\
    $M_1, M_2, r \leftarrow$ \texttt{matrix\_factorization}$(M)$\\
    $T_i \leftarrow$ \texttt{reshape}$(M_i,dims_i)$, for $i=1,2$\\
    \If{$T_1.order \geq min\_rec\_ord$}{
        $core_1, factor_1 \leftarrow \texttt{recursive\_tucker}(T_1)$\\
    }
    \Else{
        $core_1, factor_1 \leftarrow \texttt{iterative\_tucker}(T_1)$\\
    }
    
    \If{$T_2.order \geq min\_rec\_ord$}{
        $core_2, factor_2 \leftarrow \texttt{recursive\_tucker}(T_2)$\\
    }
    \Else{
        $core_2, factor_2 \leftarrow \texttt{iterative\_tucker}(T_2)$\\
    }
    
    $core \leftarrow contract(core_1, core_2)$\\
    $factors \leftarrow factor_1 + factor_2$\\
    \Return{$core, factors$}\\
\end{algorithm}
\end{minipage}~
\begin{minipage}{0.49\textwidth}
\begin{algorithm}[H]
    \caption{\texttt{hierarchical\_tucker}}
    \label{algo:hierarchical_tucker_factorization}
    \SetKwInOut{Input}{input}
    \SetKwFor{Loop}{repeat}{}{end}
    \Input{tensor $T$, rank $q$}
    $HT \leftarrow \{\}$ (empty tree)\\
    $split\_point \leftarrow \lfloor T.order/2 \rfloor$\\
    $d_1,d_2,dims_1,dims_2, split\_point \leftarrow \texttt{split}(T)$\\
    $M \leftarrow$ \texttt{unfold}$(T, (d1, d2))$\\
    $M_1, M_2, r_1 \leftarrow$ \texttt{matrix\_factorization}$(M)$\\
    \If{$\texttt{length}(dims_1) > 1$}{
        $T_1 \leftarrow$ \texttt{reshape}$(M_1,[dims_1, r_1])$\\
        $M_1 \leftarrow$ \texttt{hierarchical\_tucker}$(T_1, r_1])$\\
    }
    $HT.child_1 \leftarrow M_1$\\
    Reshape $M_2$ as in eq.~\eqref{eq:step2}\\
    $M_{21}, M_{22}, r_2 \leftarrow$ \texttt{matrix\_factorization}$(M_2)$\\
    $HT.core \leftarrow$ \texttt{reshape}$(M_{21}, q, r_1, r_2)$\\
    \If{$\texttt{length}(dims_2) > 1$}{
        $T_{22} \leftarrow$ \texttt{reshape}$(M_{22}, [dims_2, r_2])$\\
        $M_{22} \leftarrow$ \texttt{hierarchical\_tucker}$(T_{22}, r_2)$
    }
    $HT.child_2 \leftarrow M_{22}$\\
    \Return $HT$
\end{algorithm}
\end{minipage}
\section{Implementation on D-Wave}
\label{sec:implementation}
This section presents details on how we utilize D-Wave in our experiments (Section~\ref{sec:dwave}), and how we solve multiple column factorization in parallel on the quantum annealer (Section~\ref{sec:parallelism}).

\subsection{Quantum annealing parameters}
\label{sec:dwave}
Each of the algorithms presented in Sections~\ref{sec:tensor_train} to \ref{sec:hierarchical_tucker} reduces the problem of computing a tensor network to the one of a binary matrix factorization, an NP-hard task that can be expressed as a QUBO (see Section~\ref{sec:matrix_factorization}). To solve that QUBO, we map its coefficients onto the quantum chip of the D-Wave 2000Q annealer, and set a number of quantum parameters such as the annealing time or the number of anneals.

Since quantum technology is noisy, the results obtained with  D-Wave 2000Q are not deterministic. Therefore, up to several thousand anneals are usually performed, and the best solution (i.e., the one yielding the lowest QUBO value) is chosen, after annealing, from the set of obtained bitstrings. The minor-embedding process relies on constructing chains of physical qubits to represent logical variable states; however those chains might disagree on the logical variable values (we call these instances chain breaks \cite{10.1088/2058-9565/ac26d2}). In these instances we need a method to either resolve broken chains or discard anneals with broken chains. We use the following annealing parameters:
\begin{enumerate}
    \item \textit{annealing time}:  set to 1 microsecond;
    \item \textit{number of anneals}: varies according to the rank of the problem being solved: for rank $r \in \{2,3,4,5,6,7,8\}$ we use the number of  anneals $\{100, 200, 400, 600, 800, 1000, 3000\}$;
    \item \textit{chain strength}: calculated using the uniform torque compensation function \cite{dwave_github} with a prefactor of 1.5.
    \item Chain break resolution is done using the majority vote function \cite{10.1088/2058-9565/ac26d2}, where the most common state in the chain of measured qubits in a given anneal is used as the logical variable value for that solution. 
    \item Everything else was set to default. Additionally, the parallel embedded QUBO coefficients were not normalized with respect to each other, which is a reasonable choice to make because all of the rank-3 QUBO's are similar to each other. Note however that for more heterogeneous problems it would make sense to normalize the QUBO coefficients with respect to each other. 
\end{enumerate}
These parameters values were determined empirically in order to obtain best annealing results over the set of experiments we present in Section~\ref{sec:experiments}. All other annealing parameters are kept at their default values.

The density and the size of the QUBOs generated from the HUBO to QUBO conversion process \cite{dwave_github} depend heavily on the elements of column factorization subproblem represented by the HUBO. We are limited by the quantum annealing hardware (specifically the LANL D-Wave 2000Q) to a minor-embedded complete graph of size 65. Empirically, this corresponds to a maximum possible rank of 8 for arbitrary QUBO connectivity. However, it is possible to factor tensors with higher rank if the QUBO sub-problems are sufficiently sparse. In order to offer a comparison across all ranks, we limit the rank to 8 in our experiments and use a complete 65 node embedding for all rank comparison experiments in Section~\ref{sec:experiments}. The complete 65 node embedding (in addition to all discrete embeddings outlined in the following Section~\ref{sec:parallelism}) was computed using a single call to \textit{minorminer} \cite{cai2014practical, dwave_github} using default parameters. Importantly, using a fixed embedding means that the high computational cost of minor-embedding is only incurred once (as opposed to repeatedly computing a minor-embedding). 

\setlength{\belowcaptionskip}{-15pt}
\begin{wrapfigure}{r}{0.5\textwidth}
    \centering
    \vspace{-2em}
    \includegraphics[width=0.49\textwidth]{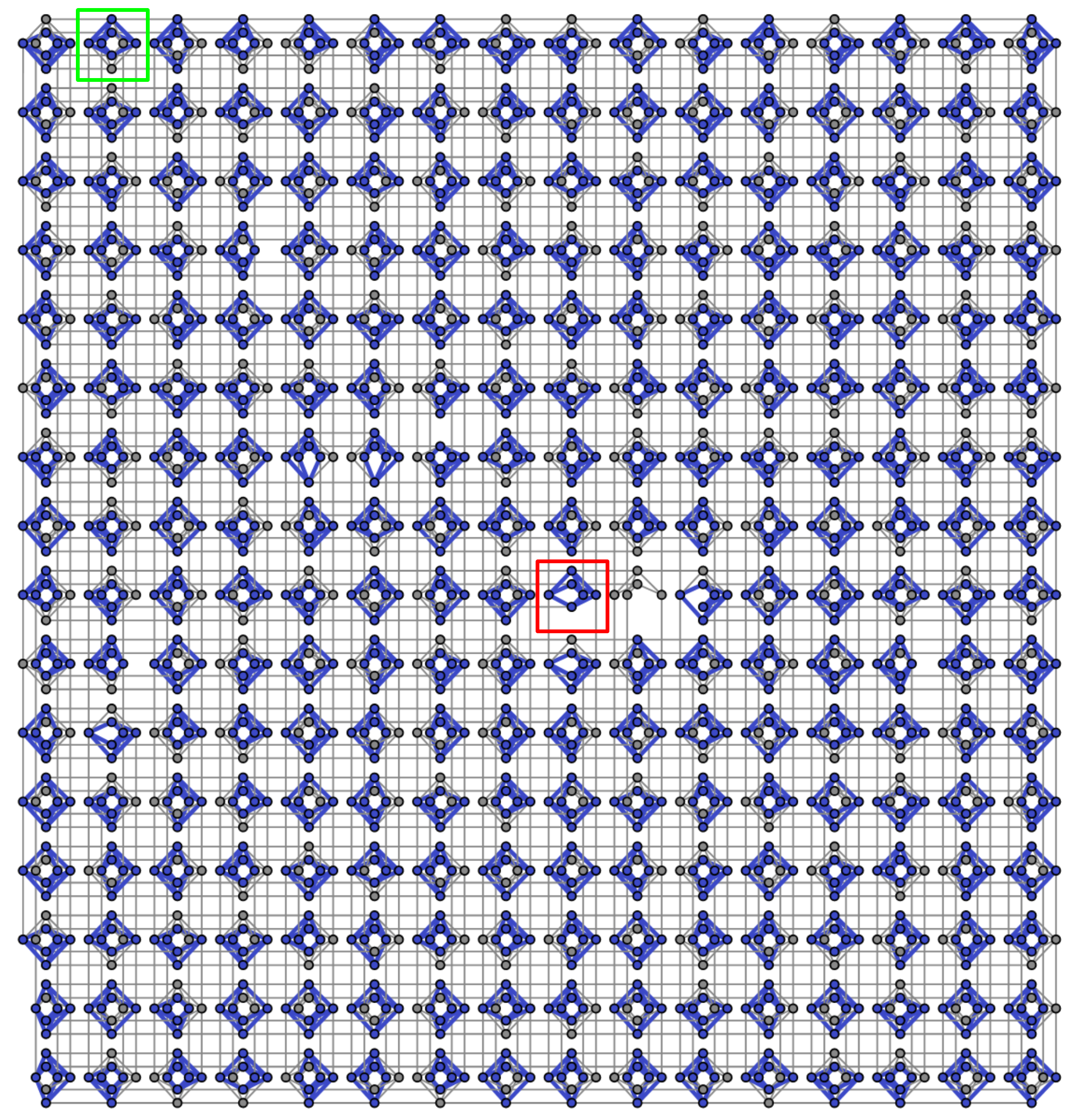}
    \caption{The Chimera graph for the LANL D-Wave 2000Q machine with 256 unit cells. Complete unit cell in green, and incomplete unit cell in red. Blue coloring on the connectivity graph shows the qubits and couplers that are used in the $255$ disjoint clique-4 minor-embeddings. }
    \label{fig:rank_3_parallel_embeddings}
\end{wrapfigure}

Finally, the D-Wave API for HUBO to QUBO conversion \cite{dwave_github}, requires the specification of a penalty factor (called \textit{strength} in the D-Wave documentation) used to rewrite higher order polynomial terms as quadratic ones. This is necessary in order to ensure the ground state solutions of the HUBO are consistent with the ground state solutions of the corresponding QUBO. This \emph{strength} parameter was always chosen as the maximum of the absolute value of any HUBO coefficient. This is a heuristic choice, which nevertheless yielded the ground state solutions of the QUBOs we solved in Section~\ref{sec:experiments}.

\subsection{Parallel quantum annealing}
\label{sec:parallelism}

The column factorization problems generated in Section~\ref{sec:matrix_factorization} are small enough to be solved on only one of the so-called Chimera unit cells of the D-Wave 2000Q quantum annealer, meaning that we can solve several column factorizations in parallel. The idea of solving problems of the type of eq.~\eqref{eq:hamiltonian} simultaneously on the D-Wave chip in one anneal has already been introduced in the literature \cite{Pelofske2022}.

Briefly, any rank-3 column factorization QUBO generated in Section~\ref{sec:matrix_factorization} will form a maximal clique of size $4$. The corresponding QUBO has $4$ linear terms, as well as some of the (at most $16$) quadratic terms. Each QUBO is solved on D-Wave 2000Q by mapping it onto the quantum hardware. The chip of the D-Wave 2000Q situated at Los Alamos National Laboratory contains $2038$ working hardware qubits, arranged in a lattice of $256$ $K_{4,4}$ bipartite graphs. The expected number of working qubits for this size of Chimera graph is $2048$; the lower number of working qubit is due to hardware defects. This hardware graph can be seen in Figure~\ref{fig:rank_3_parallel_embeddings}. Each bipartite graph is called a unit cell, and contains $16$ densely connected hardware qubits. The cells themselves are sparsely connected. Due to calibration and manufacturing defects, some of the unit cells of the D-Wave 2000Q device at Los Alamos National Laboratory contain less than $16$ qubits (for instance, the green and red squares in Figure~\ref{fig:rank_3_parallel_embeddings} show a complete and an incomplete unit cell, respectively). Importantly, each QUBO occurring in Section~\ref{sec:matrix_factorization} can be embedded onto one of the unit cells alone (with the exception of one unit cell which contains too many missing qubits to create an embedding), meaning we can solve up to $255$ column factorization problems (of rank-3) simultaneously in a single D-Wave call.

In Section~\ref{sec:experiments} we use the idea of parallel quantum annealing for the experiments looking at tensor order and tensor dimension size (these experiments use a decomposition rank of 3). For all rank comparisons we employ a fixed embedding of a complete $65$ node graph. If there is a particular matrix factorization problem with less than $255$ column factorization QUBOs, that D-Wave backend call will only make use of that number of sub-problems, not the full $255$ sub problem embedding. Each use of the $255$ sub problem embeddings first employs a random shuffle of the assigned embedding to problems in order to reduce the effect of persistent hardware biases.

\section{Experimental results}
\label{sec:experiments}
This section presents our experimental results on randomly generated tensors. We investigate the scaling in both runtime (QPU time in the case of the quantum annealer, and process CPU time for the classical case) required to solve the generated QUBOs when solving the matrix factorization sub-problems, and error rate (defined as the average number of Boolean mismatches between the input tensor and its proposed factorization, divided by the total number of tensor elements) in two scenarios: once for each original input tensor, and once for a noisy version which is obtained by flipping each bit in the input tensor independently with probability $0.001$. With this level of noise, the number of bits to be flipped can vary from zero for the smallest tensors (in which case we intentionally flip one bit at random), to around one thousand bits for the largest tensors considered in our experiments.

\renewcommand{\floatpagefraction}{.8}%
\begin{figure}
    \centering
    \includegraphics[width=0.49\textwidth]{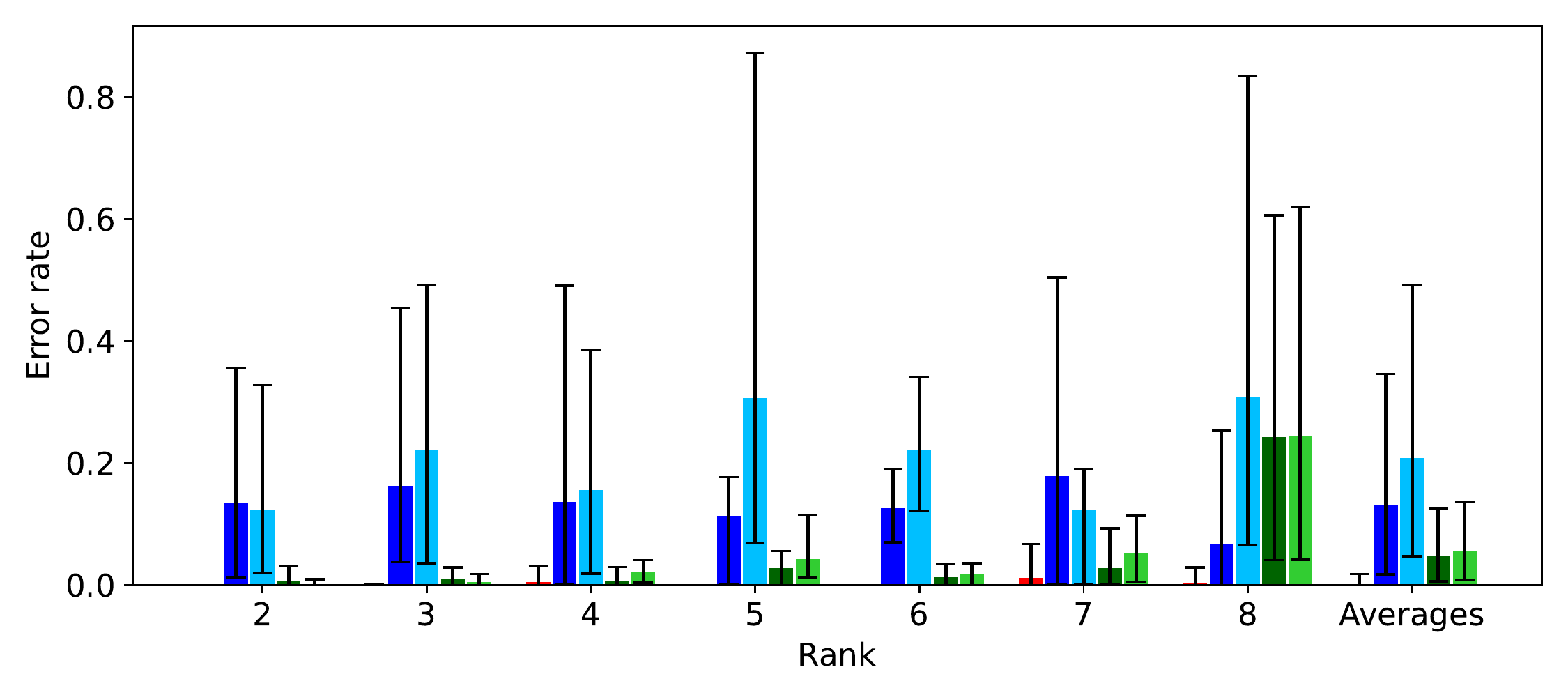}
    \includegraphics[width=0.49\textwidth]{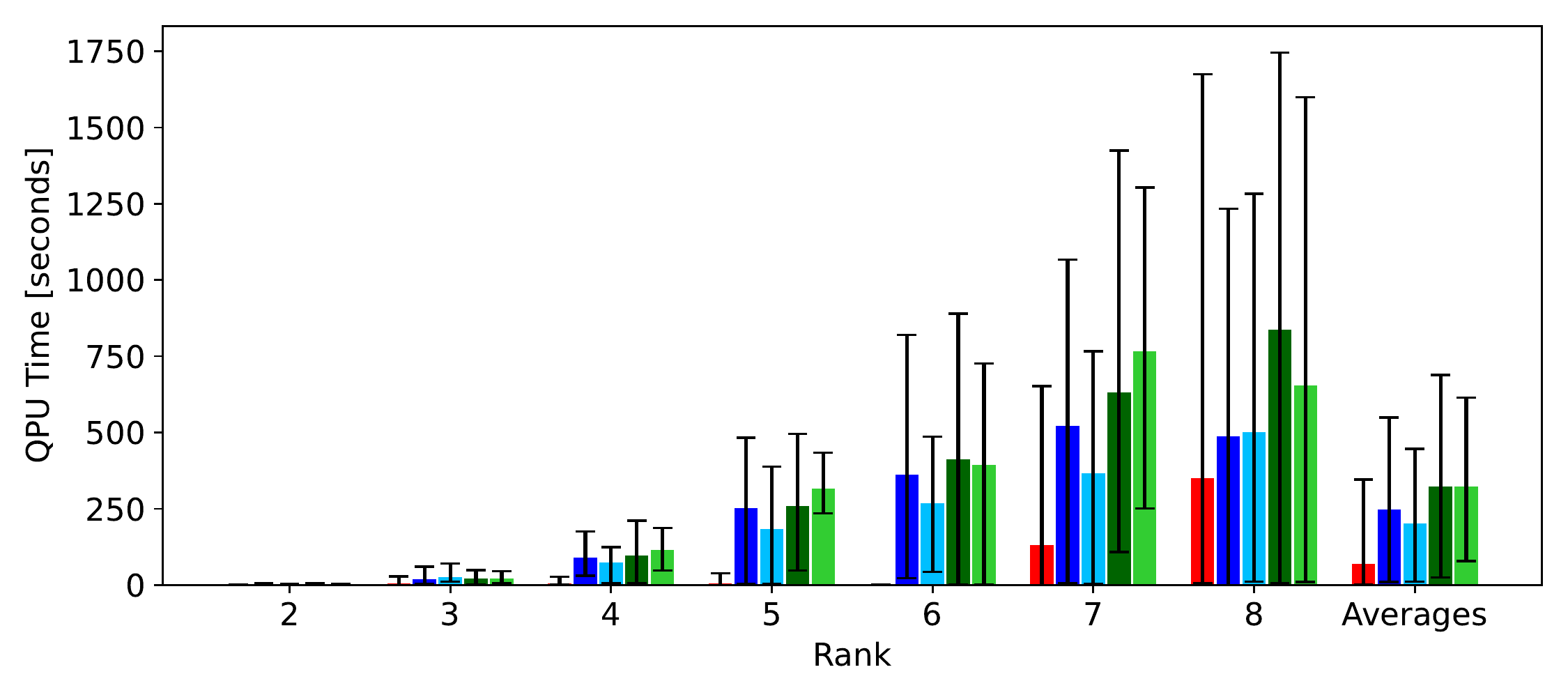}\\
    \includegraphics[width=0.49\textwidth]{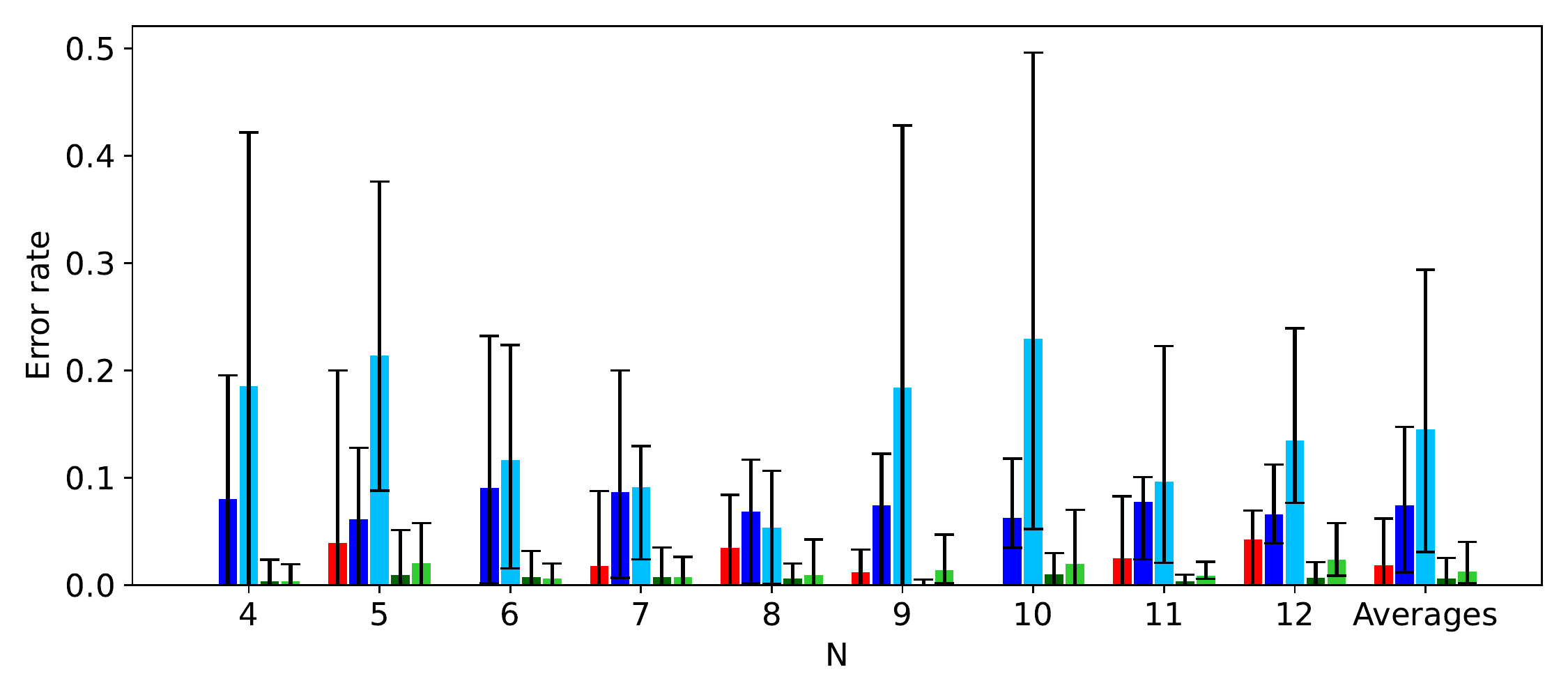}
    \includegraphics[width=0.49\textwidth]{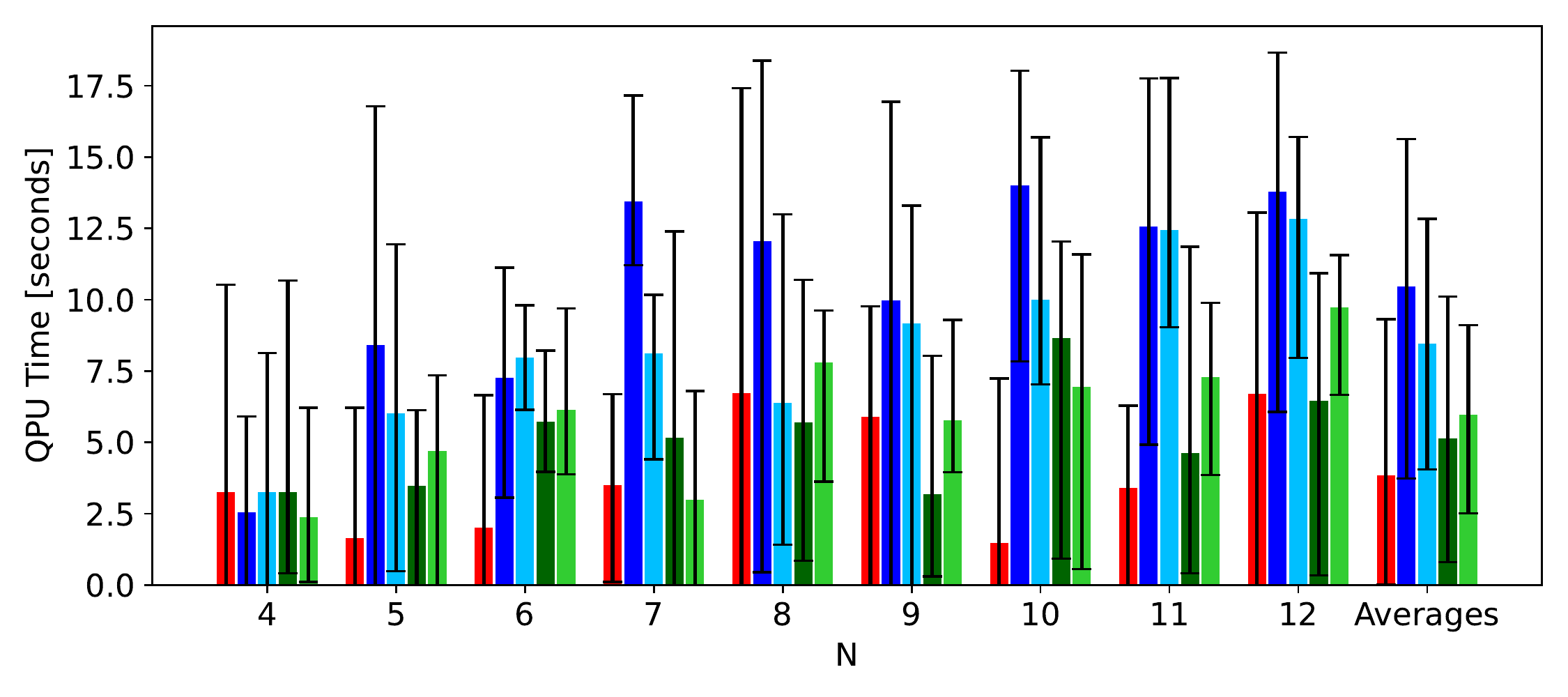}\\
    \includegraphics[width=0.49\textwidth]{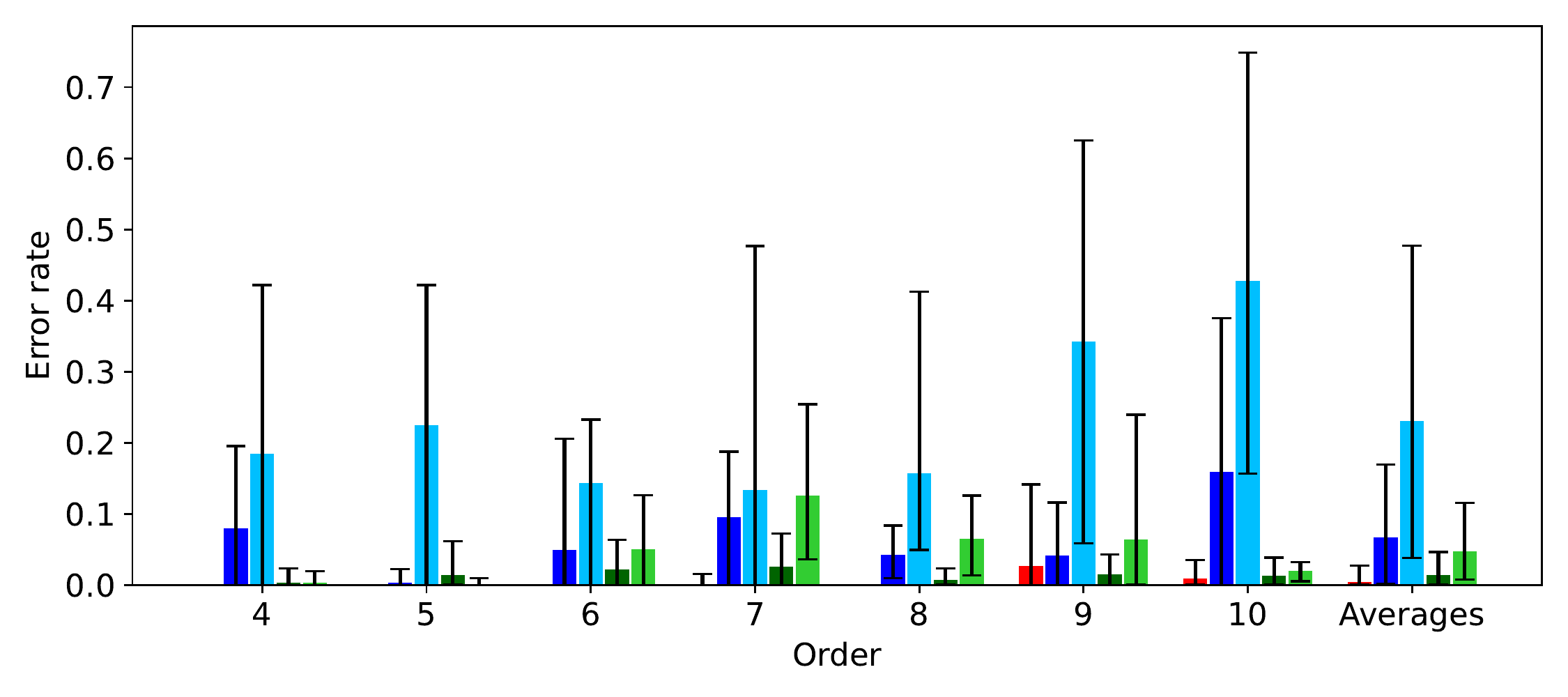}
    \includegraphics[width=0.49\textwidth]{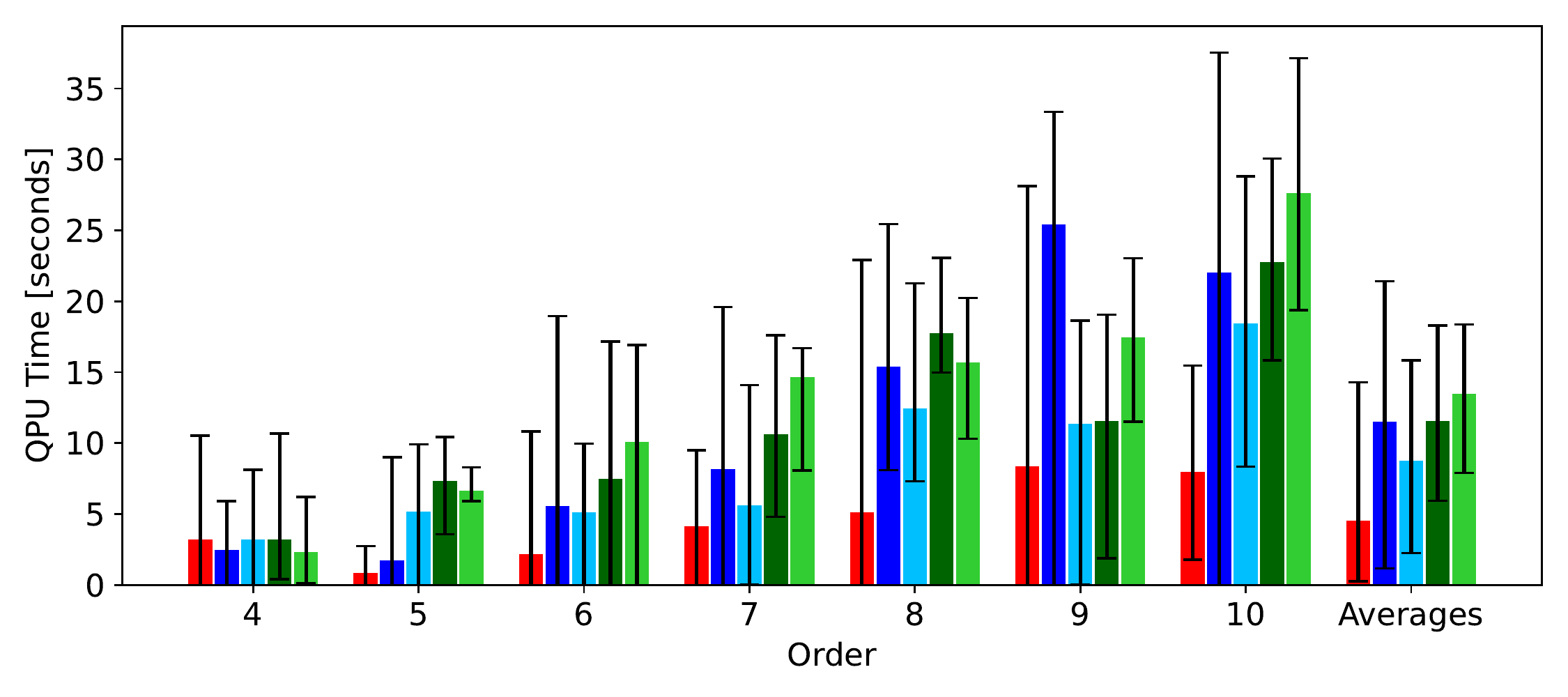}\\
    \includegraphics[width=0.49\textwidth]{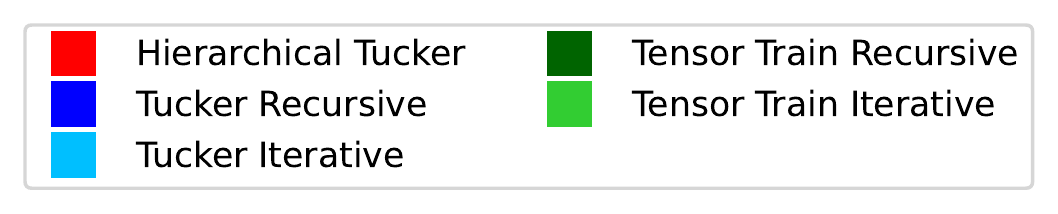}
    \caption{Tensors with no noise. Mean error rate (left column) and mean QPU time (right column). Tensor rank (top row), dimension size $N$ (middle row), and tensor order (bottom row). The right most entry on the x-axis shows the average across all previous quantities for each of the five tensor methods. Error bars indicate minimum and maximum values. }
    \label{fig:all_results_no_noise}
\end{figure}

To generate the tensors, we first generate a random tensor network of the given type (e.g., Tensor Train or Tucker), and then compute the tensor that it represents, which serves as input to our algorithms. For each network, each factor tensor or matrix is generated by sampling its binary entries from a Bernoulli distribution with probability $p$ (that is, entry $1$ with probability $p$, and entry $0$ with probability $1-p$), where $p$ is uniformly chosen in $[0.01,0.99]$. For each of the three \emph{types} of tensor algorithms, and for each combination of tensor parameters we investigate (order, dimension size, rank), we generate five different tensors, and then run the respective algorithms on those tensors. In total, we generated $330$ tensors without noise, and then added in noise to create a corresponding $330$ tensors with noise. All of our tensor algorithms can be used for multi-rank factorization, as well as varying dimension sizes. However for simplicity, we restrict both the dimension sizes and the factorization ranks to be the same.

\begin{figure}
    \centering
    \includegraphics[width=0.49\textwidth]{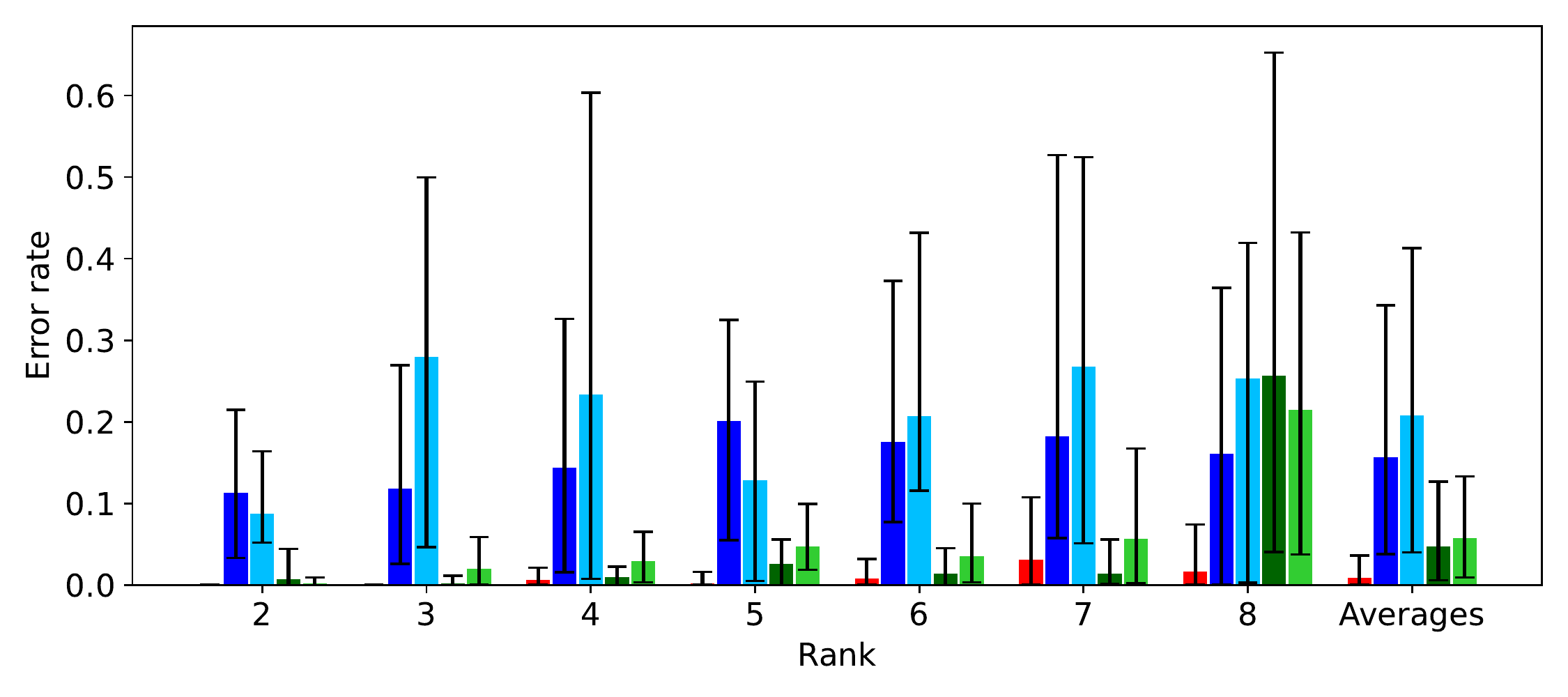}
    \includegraphics[width=0.49\textwidth]{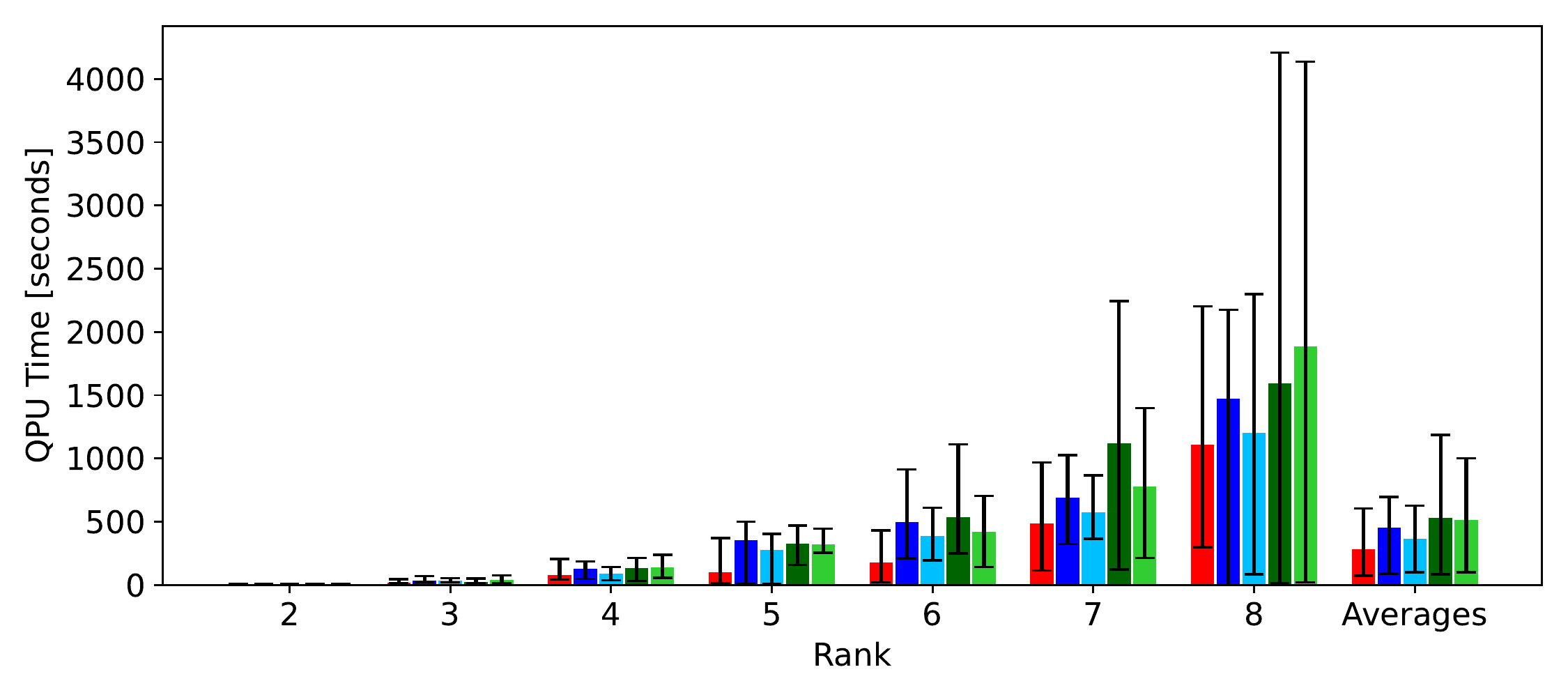}\\
    \includegraphics[width=0.49\textwidth]{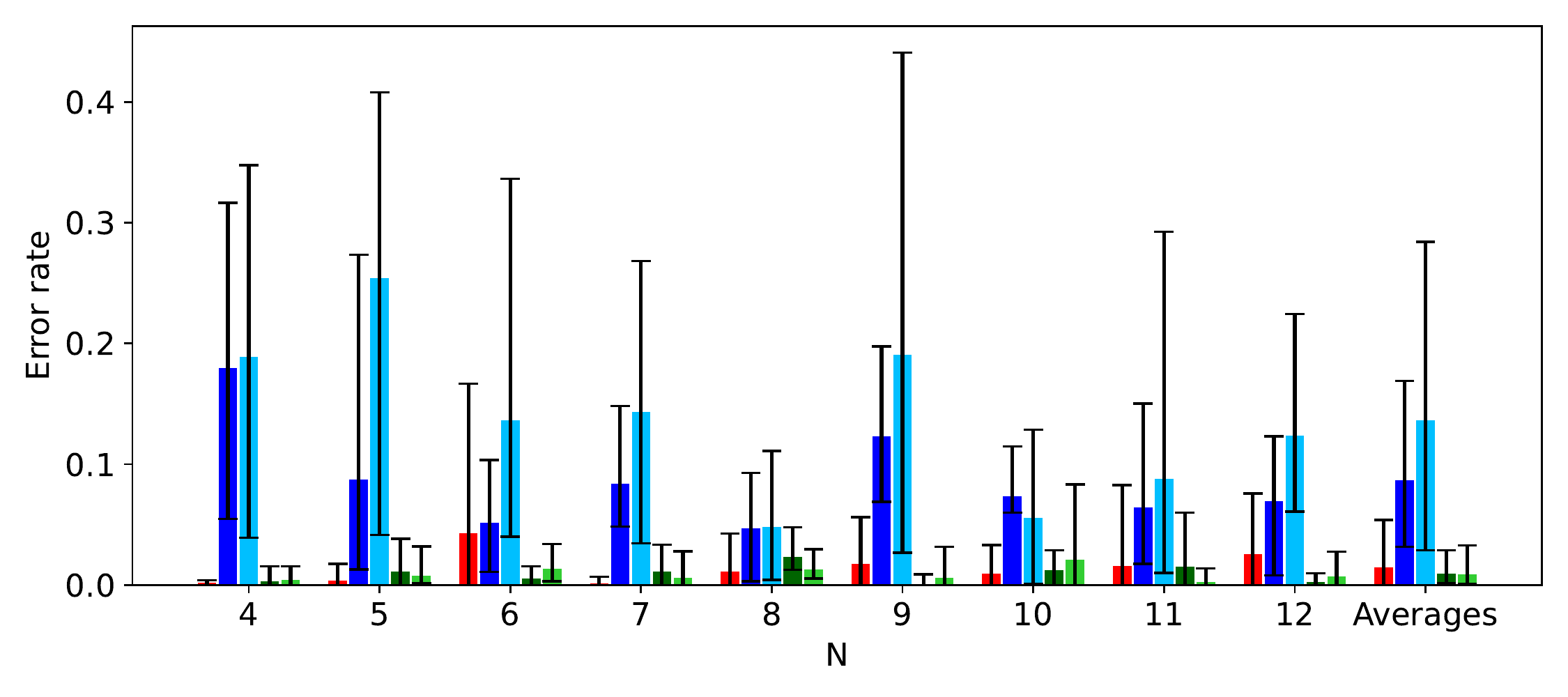}
    \includegraphics[width=0.49\textwidth]{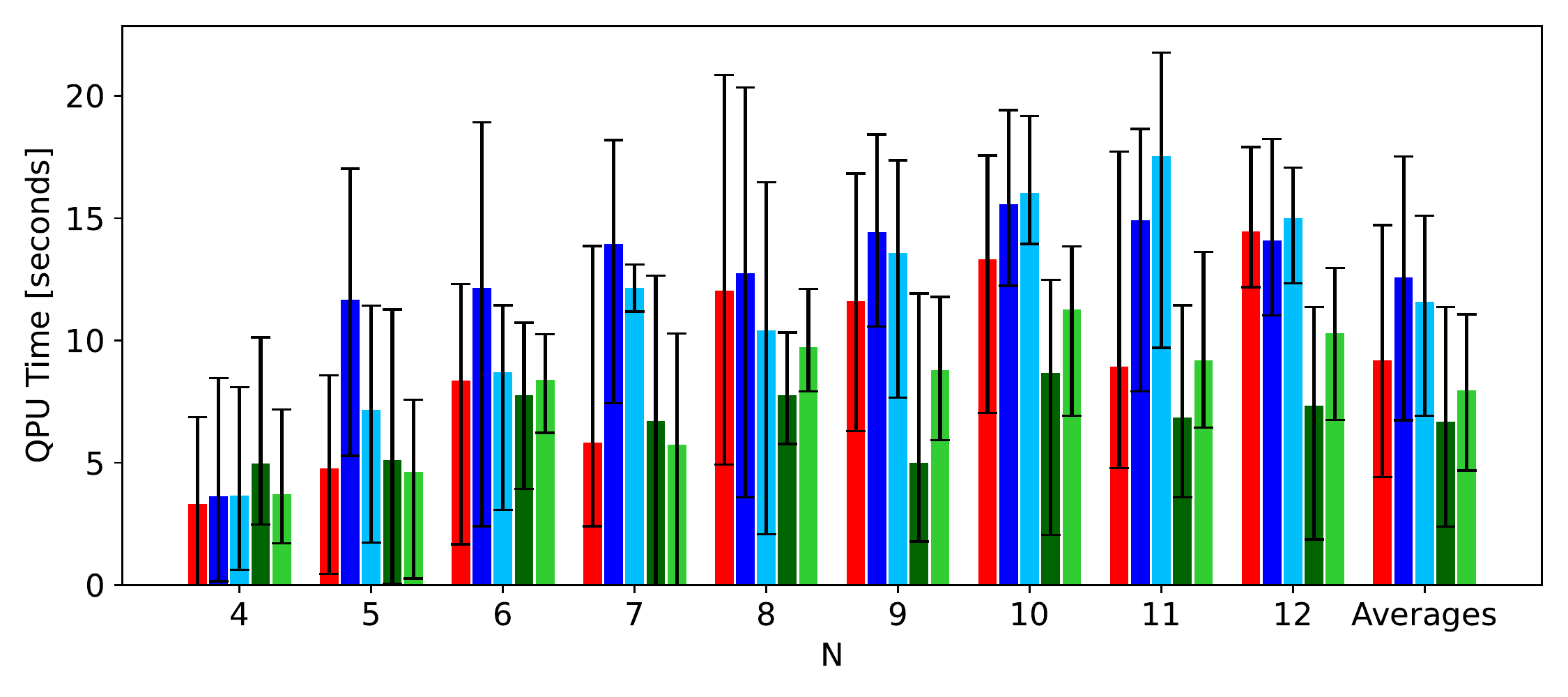}\\
    \includegraphics[width=0.49\textwidth]{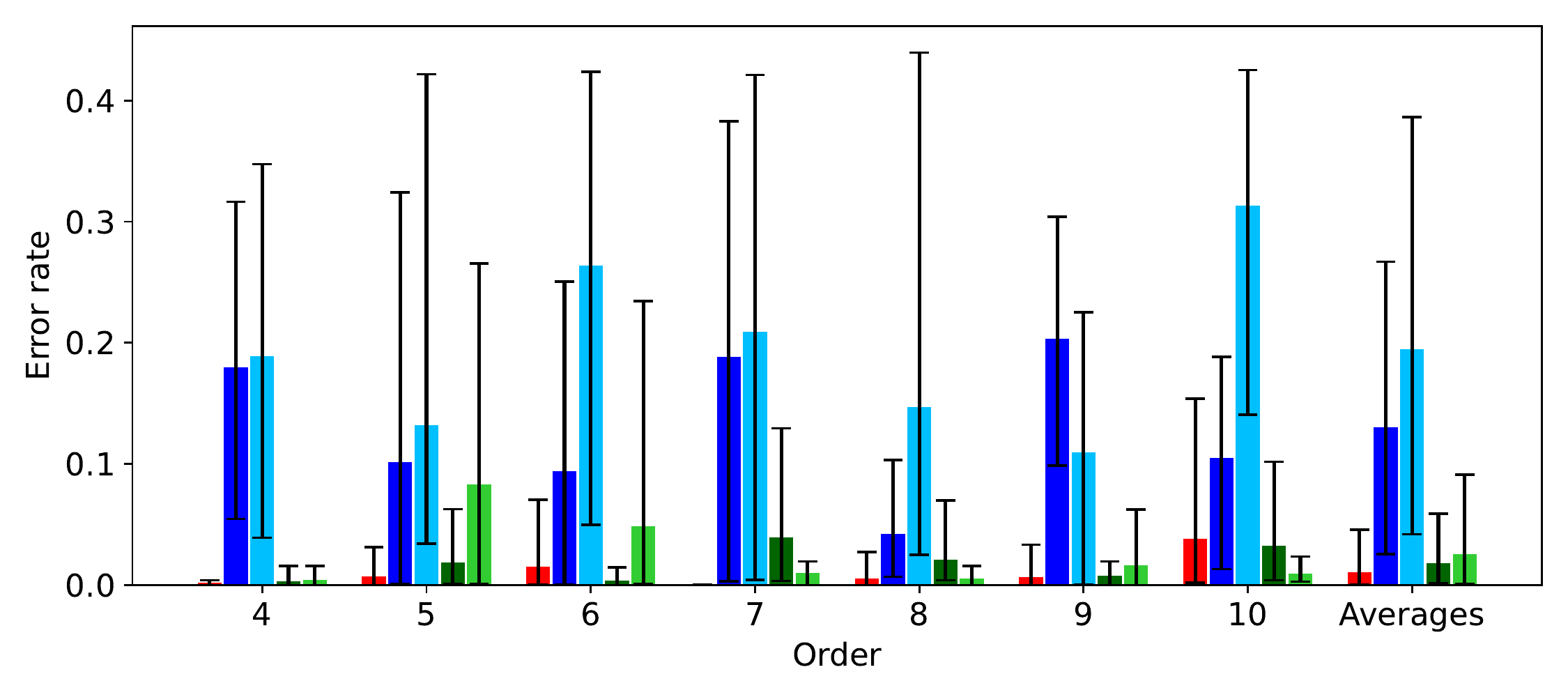}
    \includegraphics[width=0.49\textwidth]{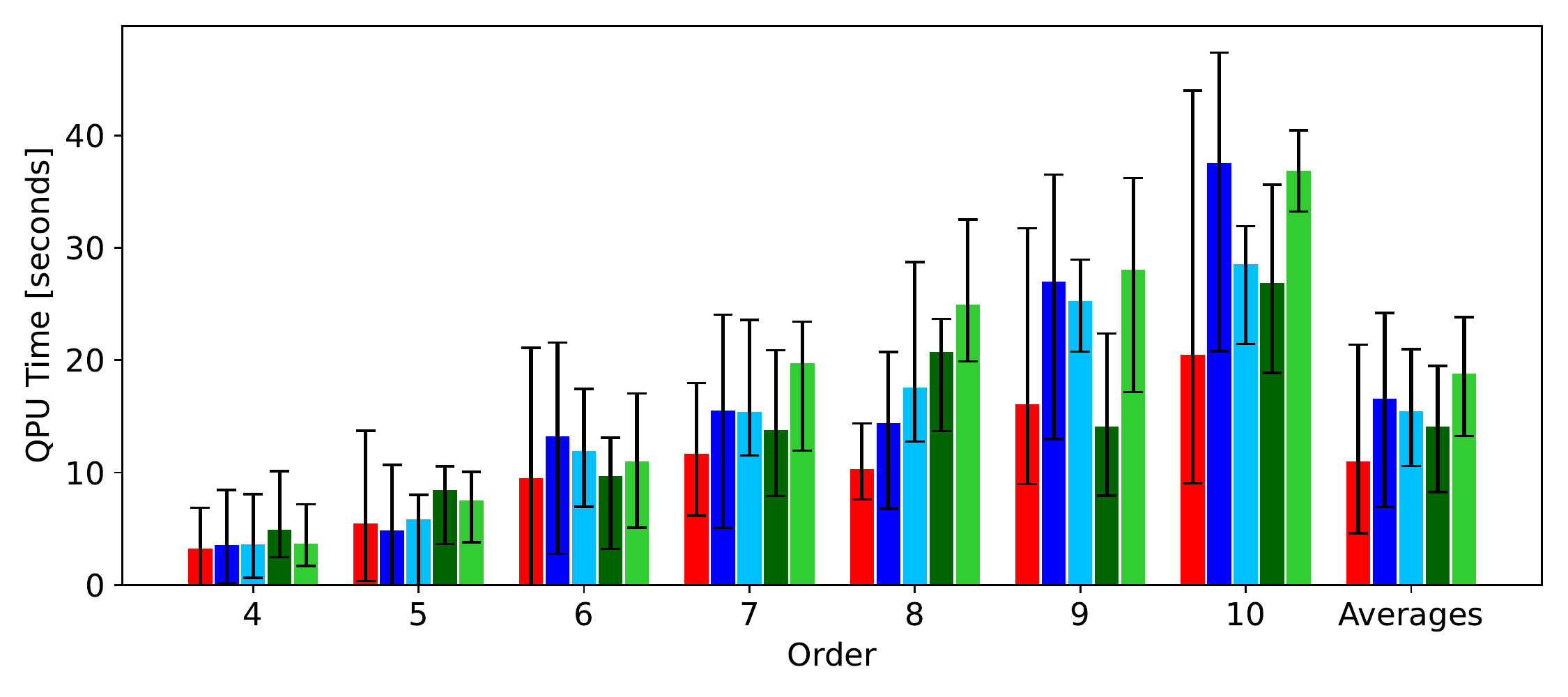}\\
    \includegraphics[width=0.49\textwidth]{figures/barplots/legend.pdf}
    \caption{Tensors with noise. Mean error rate (left column) and mean QPU time (right column). Tensor rank (top row), dimension size $N$ (middle row), and tensor order (bottom row). The right most entry on the x-axis shows the average across all previous quantities for each of the five tensor methods. Error bars indicate minimum and maximum values.}
    \label{fig:all_results_with_noise}
\end{figure}

The following sections investigate the behavior of the (iterative and recursive) Tensor Train, (iterative and recursive) Tucker, as well as Hierarchical Tucker algorithms as a function of the rank (Section~\ref{sec:results_rank}), dimension size (Section~\ref{sec:results_size}), and order (Section~\ref{sec:results_order}) of the input tensor. We evaluate all algorithms with respect to both error rate and runtime (QPU or CPU time to solve the QUBO sub-problems). In particular, the computation time we report does not include the processing steps leading up to solving the QUBO (in either the classical or quantum annealing case). For example, the time to convert the HUBO into a QUBO, or the unembedding time, is not reported in these plots. Instead we specifically investigate the scaling behavior of the required time to solve the QUBO sub-problems. We define the error rate to be the Hamming distance between $T$ (the original input tensor) and $T'$ (the reconstructed tensor from the found factors) divided by the total number of elements in $T$. Thus, an error rate of 0 means the algorithm found an exact factorization of $T$. We report the \emph{mean} metric (error rate or computation time) across the 5 test tensors for each scenario. Note that, because the tensors were generated by the type of tensor algorithm, only the recursive and iterative versions for the same tensor network type (Tucker or Tensor Train) are directly comparable to each other in the following sections (Sections~\ref{sec:results_rank}, \ref{sec:results_size}, \ref{sec:results_order}). For example, Hierarchical Tucker and Tucker results used not only different initial tensors, but also different input constructions, meaning that those results are not directly comparable (e.g., the amount of information content can be vastly different for tensors corresponding to different network types). On the other hand, both the iterative and recursive versions of Tensor Train used exactly the same tensor input structure and the exact same 5 tensors, therefore those results are directly comparable. The experimental section concludes with a comparison to the classical simulated annealing algorithm in Section~\ref{sec:classical}.

\subsection{Rank}
\label{sec:results_rank}
We start with an assessment of the accuracy as a function of the rank, while keeping the order $4$ and the dimension size $4$ fixed. Figure~\ref{fig:all_results_no_noise} (top left) and Figure~\ref{fig:all_results_with_noise} (top left) show results for all five methods under investigation for the scenario without and with added noise, respectively. Three observations are noteworthy. First, throughout all algorithms considered there does not seem to be an obvious dependence of the error on the rank of the tensor. Second, the recursive versions of both Tucker and Tensor Train result in lower error rates compared to the iterative versions. The likely reason for this finding is that a recursive version produces a tensor network of a lower depth. Since each decomposition during the construction of a network adds error, that error accumulates, and may get quite large at the leaves of the network. Therefore, it seems sensible that networks of a lower diameter should in general produce better approximations of the original tensor. Third, when adding noise, the error rate for most methods increases (the primary exception being Tensor Train Iterative).

Similarly to the error rate comparison, Figure~\ref{fig:all_results_no_noise} (top right) and Figure~\ref{fig:all_results_with_noise} (top right) investigate the scaling in QPU time as a function of the rank while keeping the order $4$ and the dimension size $4$ fixed, again for the scenario without and with added noise, respectively. The runtime increases for all methods as the rank increases, which is to be expected because as the rank increases we also increase the number of samples. When directly comparing the recursive and iterative versions of Tensor Train and Tucker, we see that the recursive version uses more than or equal to the iterative version. Lastly, adding noise to the tensor makes the decomposition more difficult, expressed in a higher QPU time throughout all methods and ranks.

Averages for both error rates and QPU times confirm that the recursive versions are more accurate than the iterative ones while being roughly equally fast.

\subsection{Size in each dimension}
\label{sec:results_size}
Similarly to Section~\ref{sec:results_rank}, we investigate the scaling of both error rate and QPU runtime as a function of the dimension size of the input tensor. Figure~\ref{fig:all_results_no_noise} (middle left) and Figure~\ref{fig:all_results_with_noise} (middle left) shows the mean error rate results of this experiment for the scenario without and with added noise, respectively. As observed in Section~\ref{sec:results_rank}, there is no obvious dependence of the error on the dimension size, the recursive version of both Tucker and Tensor Train give a lower error rate than the iterative version, and adding noise to the tensor decreases the accuracy throughout all methods, as expected. This can also been seen by looking at the averages over all sizes.

The mean QPU runtime as a function of the dimension size, reported in Figure~\ref{fig:all_results_no_noise} (middle right) for the scenario without noise and in Figure~\ref{fig:all_results_with_noise} (middle right) for the scenario with added noise, shows a (weak, possibly linear) dependence on the dimension size, where noisy tensors again consistently require a higher runtime and have larger error bars.

\subsection{Order}
\label{sec:results_order}
Last, we investigate the scaling of error rates and QPU times as a function of the order of the tensor while keeping the factorization rank $3$ and the tensor dimension size $4$ fixed. Results are displayed in Figure~\ref{fig:all_results_no_noise} (bottom left) and Figure~\ref{fig:all_results_with_noise} (bottom left) for the scenario without and with added noise, respectively. We observe a similar picture as for the previous experiments, with no obvious dependence of the error on the order. We again observe that the recursive versions of Tucker and Tensor Train result in lower error rates than their iterative counterparts, which is also reflected in the averages across all orders.

The mean QPU time for the scaling in the order of the input tensor is given in Figure~\ref{fig:all_results_no_noise} (bottom left) for the scenario without noise and in Figure~\ref{fig:all_results_with_noise} (bottom left) for the scenario with added noise. We observe that the QPU scaling behaves very similarly to the scaling in the dimension size of the tensor, exhibiting a seemingly (weak, linear) increase. Importantly, QPU times for both the scaling in the tensor size and order are in the vicinity of seconds, demonstrating that tensor decomposition with the help of quantum annealing is feasible in practice.

\subsection{Classical algorithm comparison}
\label{sec:classical}
We repeat the comparison of the Hierarchical Tucker, Tensor Train, and Tucker decomposition algorithms using two classical heuristic methods to solve the QUBO's generated by the matrix factorization sub-routine, instead of the D-Wave quantum annealer. The first classical heuristic we use is the implementation of simulated annealing provided by D-Wave Systems, Inc., available at \url{https://github.com/dwavesystems/dwave-neal} with all default settings (except for the number of samples). The second classical heurisitc we use is a greedy steepest descent algorithm, also available on Github at \url{https://github.com/dwavesystems/dwave-greedy}. For a fair comparison, we use the same number of samples as with the quantum annealer (for rank 3 this was $200$ samples).

We use the experimental setting introduced in Section~\ref{sec:experiments}. As before, we report error rates and either CPU or QPU times (in particular, the \textit{qpu-access-time} for the QA backend solving the QUBO sub-problems, and the \textit{cpu-process} time for the classical solvers) depending on whether we look at a quantum or classical implementation. We copy the setting of Section~\ref{sec:results_size}, though we only consider the largest tensor dimension size 12 therein and keep the order 4 and the rank 3 fixed. As an additional comparison, we run the two classical methods using \emph{sequential} QUBO solving (this means solving each of the small QUBO's one at a time) as well as the \emph{parallel} method we use in the quantum annealing implementation (where many disjoint QUBO's are combined into a larger QUBO which is then solved as a single QUBO). We add this sequential and parallel difference to the classical methods because we expect the sequential method to be faster for classical methods; but we also expect the parallel method to be faster for quantum annealing. Therefore such a comparison is warranted.

\begin{wrapfigure}{r}{0.49\textwidth}
    \centering
    \vspace{-0.1em}
    \includegraphics[width=0.49\textwidth]{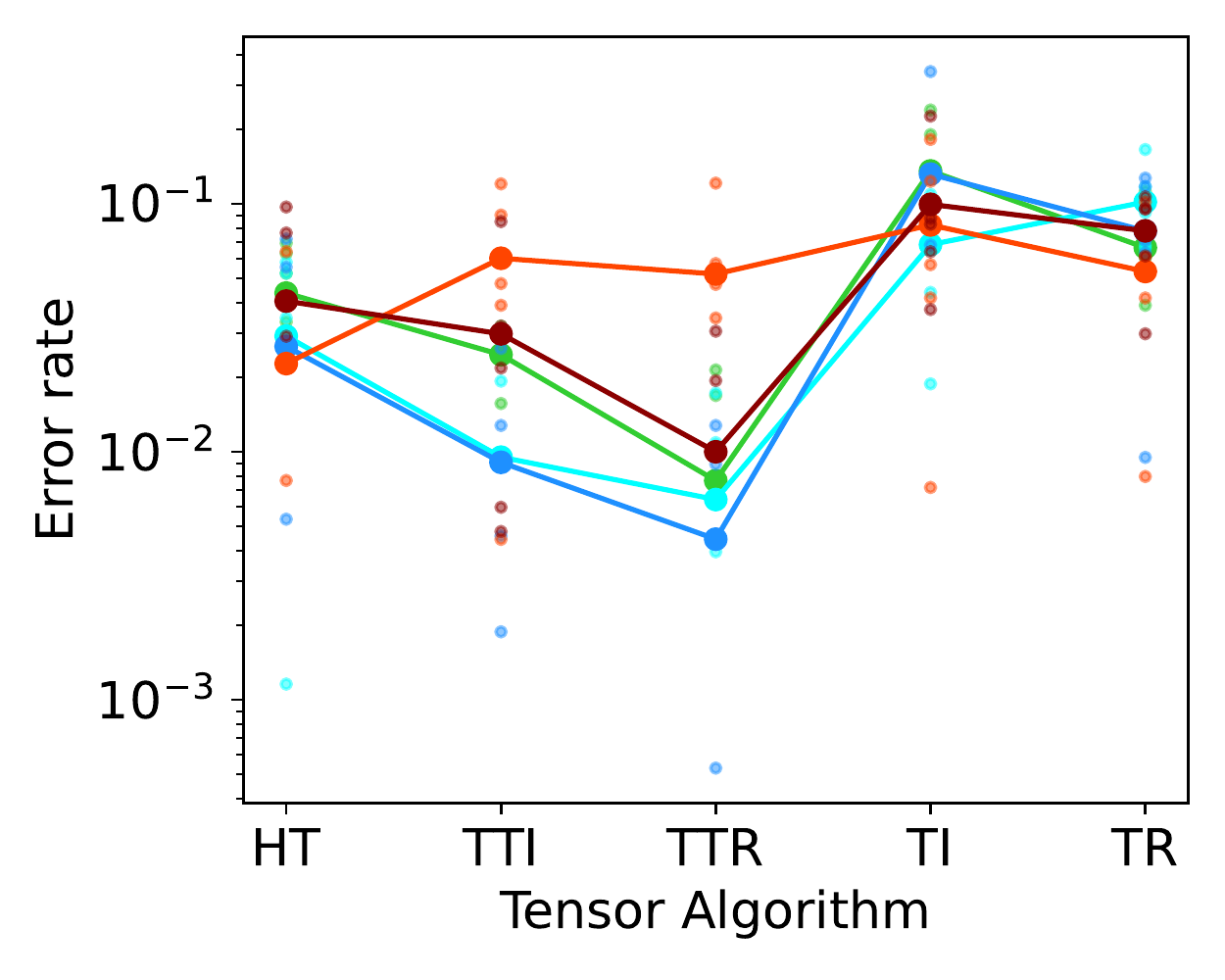}\hfill
    \includegraphics[width=0.49\textwidth]{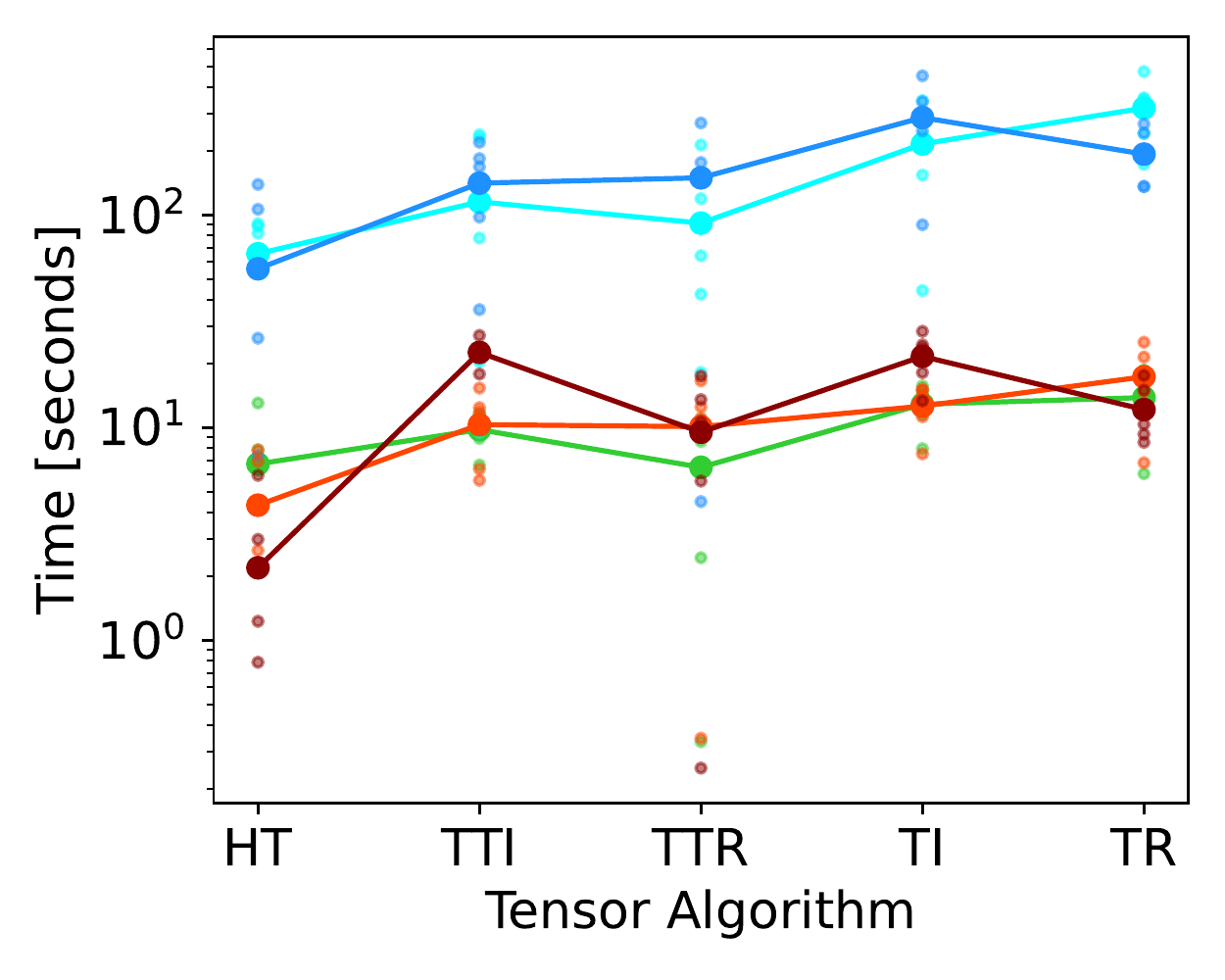}
    \includegraphics[width=0.49\textwidth]{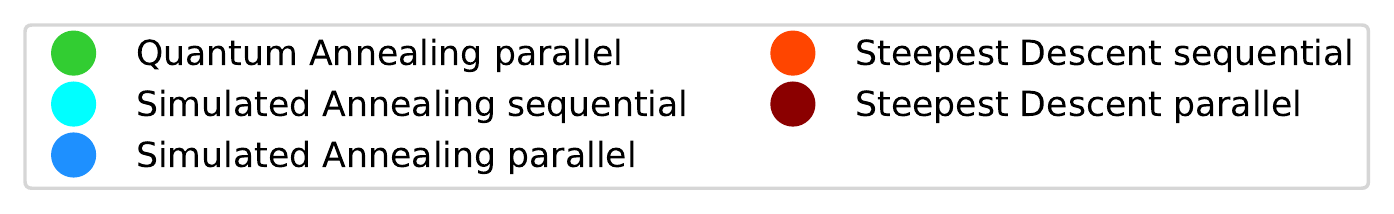}
    \caption{Classical (simulated annealing and steepest descent algorithms) and quantum annealing implementations (D-Wave 2000Q) of the five tensor algorithms Hierarchical Tucker (HT), Tensor Train Iterative (TTI), Tensor Train Recursive (TTR), Tucker Iterative (TI), and Tucker Recursive (TR). The tensor parameters are fixed to be rank $3$, dimension size $12$, and order $4$. For D-Wave we report QPU time (in particular, \textit{qpu-access-time}), whereas for classical computations we report CPU \textit{process} time. Log scale on the y-axes. 
    Each dot is one tensor (5 tensors per algorithm), while lines connect the mean values for each algorithm. \vspace{-5em}}
    \label{fig:classical_comparison}
\end{wrapfigure}

Figure~\ref{fig:classical_comparison} (left) shows that quantum annealing results in lower error rates than greedy steepest descent. Moreover, we observe that the recursive versions of the Tensor Train and Tucker algorithms result in lower error rates than their iterative counterparts.

We observe that simulated annealing (with default settings) takes significantly more computation time across all tensor algorithms than greedy steepest descent or quantum annealing (Figure~\ref{fig:classical_comparison}, right). Interestingly, quantum annealing and greedy steepest descent are comparable in terms of computation time.
\section{Conclusion}
\label{sec:discussion}

This article considers Boolean tensor networks, or the factorization of a Boolean tensor into lower dimensional tensors. At the lowest level, this task reduces to performing a large number of Boolean matrix factorizations. Boolean matrix factorization is a hard optimization problem that we solve on the D-Wave 2000Q quantum annealer, after reformulating it as quadratic unconstrained binary optimization.

We show that Boolean input tensors can be efficiently decomposed into Boolean tensor networks of a certain shape using a set of basic operations---unfolding, reshaping, and Boolean matrix factorization. The latter is solved with the help of quantum annealing. We implement those operations in three methods, called Tensor Train, Tucker, and Hierarchical Tucker algorithms, of which the recursive versions are a novel contribution of this work. We show that at the lowest level of the recursion, several of the created QUBO problems can be solved on the D-Wave 2000Q quantum annealer with the help of \textit{parallel quantum annealing} in the same backend call.

We experimentally demonstrate the viability of all three algorithms in an experimental study. On synthetically generated Boolean input tensors of varying ranks, sizes, and orders, we show that our algorithms in connection with quantum annealing allow one to accurately factor input tensors containing up to a million elements. We see that the recursive versions of the Tucker and Tensor Train algorithms consistently result in lower error rates than the iterative versions. A comparison to classical solvers, obtained by replacing the solving step of the quadratic unconstrained binary optimization via D-Wave with simulated annealing, shows that the approach involving the D-Wave 2000Q uses considerably less computation time in comparison to the classical simulated  annealing while returning comparable error rates. All of the algorithms presented in this paper are available online \cite{Pelofske2021_pyQBTNs}.

\section*{Acknowledgments}
This work has been supported by the US Department of Energy through the Los Alamos National Laboratory. Los Alamos National Laboratory is operated by Triad National Security, LLC, for the National Nuclear Security Administration of U.S. Department of Energy (Contract No.~89233218CNA000001) and by the Laboratory Directed Research and Development program of Los Alamos National Laboratory under project numbers 20190065DR and 20190020DR as well as under 20180267ER.
The work of Hristo Djidjev has been also partially supported by the Grant No.~BG05M2OP001-1.001-0003, financed by the Science and Education for Smart Growth Operational Program (2014-2020) and co-financed by the European Union through the European Structural and Investment Funds.

\section*{Data availability}
All datasets and code are available online at \url{https://github.com/lanl/pyQBTNs}.

\printbibliography
\end{document}